\documentclass[preprint,floats,aps,epsfig,nofootinbib,amssymb]{revtex4-1}
\usepackage{mathrsfs}
\usepackage{slashed}
\usepackage{graphicx,color}
\usepackage{subfigure}
\usepackage{dcolumn}
\usepackage{bm}
\usepackage{subfig}
\usepackage{graphicx}
\usepackage{mathrsfs,amsmath,amssymb}
\usepackage{stackrel}
\usepackage{bbding}
\usepackage{pgfplots}
\usetikzlibrary{positioning}
\usetikzlibrary{snakes}
\def \bea {\begin{eqnarray}}
\def \eea {\end{eqnarray}}

\begin{document}

\title{ Axion-Inflation {\it Baryogenesis} via  New U(1) gauge symmetries }

\author{Wei Chao$^{1,2}$}
\email{chaowei@bnu.edu.cn}
\author{Yonghua Wang$^{1}$}
\email{wangyh@mail.bnu.edu.cn}
\author{ChenHui Xie$^{1}$}
\email{chenhuixie@mail.bnu.edu.cn}

\affiliation{$^1$Center of Advanced Quantum Studies, School of Physics and Astronomy, Beijing Normal University, Beijing, 100875, China \\
$^2$Key Laboratory of Multi-scale Spin Physics, Ministry of Education,
Beijing Normal University, Beijing 100875, China}
\vspace{3cm}

\begin{abstract}

We investigate axion-inflation baryogenesis models, embedded into  U(1) gauge symmetric extensions of the Standard Model (SM),  in which the new gauge field couples to the pseudo-scalar inflaton via the Chern-Simons coupling.  The motion of the inflaton induces a techyonic instability for one of the two helicities of the gauge field, resulting in the production of the helical gauge field. It further leads to the generation of the SM particle number densities via anomalies during the reheating, which is sufficient to generate the matter-antimatter asymmetry of the universe. Our numerical results show that  this mechanism works for the  $U(1)_{\mathbf L}$, $U(1)_{\mathbf R}$ and $U(1)_{\mathbf{B-L}}$ gauge symmetry cases, where subscripts ${\mathbf L}$, $\mathbf{R}$, $\mathbf{B-L}$ indicate the lepton number, the right-handed fermion and the baryon number minus the lepton number respectively.  The key point for these mechanisms to work is that the evolution of the number density for right-handed neutrinos is decoupled from those of the SM particles, which shares the same merit as the Dirac Leptogenesis mechanism. 

\end{abstract}

\maketitle
\section{Introduction}

The Standard Model (SM) of particle physics~\cite{Weinberg:1967tq} has achieved great success. All particles predicted by the SM have been discovered~\cite{ATLAS:2012yve,ATLAS:2015yey}. However, the SM can only be a low energy effective field theory, as it cannot address problems of the light but non-zero active neutrino masses~\cite{Super-Kamiokande:1998kpq,SNO:2002tuh,DayaBay:2012fng},  the cold dark matter~\cite{Bertone:2004pz} and the baryon asymmetry of the universe (BAU)~\cite{Dine:2003ax}.  The BAU, which is quantified as the ratio of the baryon density to the photon density, is measured at the time of the Big Bang Nucleosynthesis (BBN) and the Cosmic Microwave Background (CMB)~\cite{Planck:2018vyg}, with
\bea
\eta = \frac{n_b }{n_\gamma} = (6.10\pm0.4) \times 10^{-10} \; .
\eea
To dynamically generate the the BAU, a barygenesis mechanism must satisfy the so-called Sakharov conditions~\cite{Sakharov:1967dj}, which are (1) the Baryon number ($\mathbf{B}$) violation; (2)the  $\mathbf{C}$ and $\mathbf{CP}$ violations; and (3) the departure from thermodynamic equilibrium.  In the SM,  $\mathbf{B}$ is violated by the sphaleron process,  and there is $\mathbf{CP}$ violations in the CKM matrix and the PMNS matrix. However no process in the SM goes out of thermal equilibrium in the early universe, which is a sign of new physics beyond the SM.  

There are many {\it Baryogenesis} models, of which representative ones include Leptogenesis~\cite{Fukugita:1986hr}, Electroweak Baryogenesis (EWBG)~\cite{Cohen:1993nk,Trodden:1998ym,Morrissey:2012db},  Afleck-Dine Baryogenesis (ADB)~\cite{Affleck:1984fy} and Spontaneous Baryogenesis~\cite{Cohen:1988kt}.  Leptogenesis mechanism addresses the BAU via the CP-asymmetric decay of heavy seesaw particles, which also explain the tiny active neutrino masses via the seesaw mechanism. EWBG relies on the first order electroweak phase transition and the CP-violating scatterings of SM particles off the expanding bubble wall generate non-zero number densities of various particles, of which number densities of left-handed fermions are biased by the electroweak sphaleron process resulting non-zero $\mathbf{B}$ density.  Once eaten by the expanding bubble, inside which the sphaleron is quenched, it contributes to the final BAU.  Although detectable, EWBG suffers from the tension between the requirement of large CP violation and the non-observation of the electric dipole moments of the electron or neutron~\cite{Li:2010ax,Chao:2017oux}.  ADB generates the BAU with the help of the dynamical evolution of the Aflick-Dine field which carries non-zero $\mathbf{B}$ or $\mathbf{L}$.  Spontaneous Baryogenesis mechanism  generates the BAU with the help of the CPT violation.
There are  also new ideas on {\it Baryogenesis} models, including axiogenesis~\cite{Co:2019wyp,Domcke:2020kcp}, axion-inflation {\it Baryogenesis} (AIB)~\cite{Domcke:2019mnd}, QCD {\it Baryogenesis}~\cite{Croon:2019ugf}, Wash-in Leptogenesis~\cite{Domcke:2020quw,Marshak:1979fm}, Lepton-flavorgenesis~\cite{Mukaida:2021sgv}, Hylogenesis~\cite{Davoudiasl:2010am}, Darkogenesis~\cite{Shelton:2010ta}, WIMP-Triggered {\it Baryogenesis}~\cite{Cui:2011ab}, freeze-in {\it Baryogenesis}~\cite{Hall:2010jx}, Mesogenesis~\cite{Elor:2018twp}, Majorogenesis~\cite{Chao:2023ojl}, etc.  We refer the reader to Ref.~\cite{Elor:2022hpa} for  descriptions of those models in detail.

In this paper, we work in the framework of the axion-inflation {\it Baryogenesis}, in which the pseudo-scalar inflaton~\cite{Freese:1990rb} couples to the Chern-Simons gravity~\cite{Alexander:2004us} or gauge fields~\cite{Akita:2017ecc,Domcke:2019mnd,Maleknejad:2020yys,Maleknejad:2020pec,Domcke:2022kfs,Hashiba:2021gmn,Tishue:2021blv,Cado:2021bia,Kamada:2019ewe}.  In this way, helical gravitational wave or gauge field will be produced towards the end of the inflation, and non-zero particle number densities can be generated by these helical GW or gauge field through the chiral anomaly. It has been shown in Refs.~\cite{Jimenez:2017cdr}  that the primordial hyper magnetic fields generated during axion inflation allows to generate the BAU  around the time of the electroweak phase transition.  One may ask what is the case if the generated helical gauge field produce chiral fermions simultaneously during the reheating,  especially when the related gauge symmetry is spontaneous broken just after the inflation. The aim of this paper is to check whether the new U(1) gauge field model can  produce  the observed BAU via the axion-inflation baryogenesis mechansim. We extend the SM with $U(1)_{\mathbf L}$, $U(1)_{\mathbf B}$~\cite{FileviezPerez:2010gw,Dulaney:2010dj,Chao:2010mp}, $U(1)_{\mathbf R}$~\cite{Langacker:2008yv,Chao:2017rwv} and $U(1)_{{\mathbf{B-L}}}$~\cite{Mohapatra:1980qe,Marshak:1979fm,Wetterich:1981bx} gauge symmetries, where the subscript $\mathbf{R}$ indicates right-handed fermions. It has been shown that both $U(1)_{\mathbf{R}}$ and $U(1)_{\mathbf{B-L}}$ are anomaly free gauge symmetries within the SM particle content plus three right-handed neutrinos, while the $U(1)_{\mathbf L}$ and $U(1)_{\mathbf B}$ symmetries require new chiral fermions for anomaly cancellations. We calculate the primordial baryon and lepton asymmetries induced by  the triangle anomalies of these gauge fields,  then simulate the final BAU by solve the complete transport equations.  Numerical results that  $U(1)_{\mathbf L}$, $U(1)_{\mathbf R}$ and $U(1)_{\mathbf{B-L}}$ may  address the  BAU,  while the $U(1)_{\mathbf B}$ can not, in the minimal framework.  The key point for this mechanism to work is that the evolution for right-handed neutrinos are decoupled from that of the SM particles, which is true whenever neutrinos are Dirac particles~\cite{Dick:1999je}. It should be mentioned that  those models might also work for case where right-handed neutrinos are heavy Majorana particles,  in which the BAU can be generated by the wash-in Leptogenesis mechanism.

The remaining of this paper is organized as follows: In section II we present various U(1) models and calculate triangle anomalies.  In section III we calculate  helical gauge fields produced by the axion-inflaton. Section IV is devoted to the calculation of the BAU induced by gauge fields in various models. The final part is concluding remarks.

\section{U(1) models and triangle anomalies}

In this section we consider new gauge symmetric extensions to the SM and the triangle anomalies associated to them. 
Considering the complexity of the new non-Abelian gauge symmetry extended models, we only focus on Abelian gauge group extended models  in this paper. 
There are many possible $U(1)$ extensions to the SM~\cite{Langacker:2008yv}, of which the gauged $\mathbf{B-L}$~\cite{Mohapatra:1980qe,Marshak:1979fm,Wetterich:1981bx}, $\mathbf{B}$, $\mathbf{L}$~\cite{FileviezPerez:2010gw,Dulaney:2010dj,Chao:2010mp}, $\mathbf{B+L}$~\cite{Chao:2015nsm,Chao:2016avy}, $\mathbf{L_i-L_j}$~\cite{He:1991qd}  have received great attentions. 
The $U(1)_{\mathbf{B-L}}$ is a minimal extension to the SM for automatic anomaly cancellations.  In addition, the $U(1)_\mathbf{R}$~\cite{Chao:2017rwv}, which is the gauge symmetry for right-handed fermions, shares the same merit as the $U(1)_\mathbf{B-L}$ on anomaly cancellations, but this model is severely constrained by the $Z-Z^\prime$ mixing. We list in the table.~\ref{table1}, charges of various SM  particles in U(1)  gauge symmetries, as well as new particles required for anomaly cancellations, where $Q_L$, $u_R$, $d_R$ are SM quarks, $\ell_L$ and $E_R$ are SM leptons,  $N_R$ is right-handed neutrino, $H$ is the SM Higgs doublet, $\psi_{L, R}$, $\eta_{L, R}$ and $\chi_{L, R}$ are  charged as $(1, 2, \frac{1}{2})$, $(1, 1, 1)$ and $(1, 1, 0)$ under the $SU(3)_C \times SU(2)_L \times U(1)_Y$ group. It is easy to check that all anomalies are cancelled  in these simple frameworks, i.e., ${\cal A}_1 (SU(3)_C^2 \otimes U(1)_X)$, ${\cal A}_2 (SU(2)_L^2 \otimes U(1)_X)$, ${\cal A}_3 (U(1)_Y^2 \otimes U(1)_X)$, ${\cal A}_4 (U(1)_Y \otimes U(1)^2_X)$, ${\cal A}_5 ( U(1)_X^3)$, ${\cal A}_6 (U(1)_X)$~\cite{Witten:1982fp,Adler:1969gk,Bell:1969ts,Bardeen:1969md,Eguchi:1976db,Alvarez-Gaume:1983ihn}.  We refer the reader to Refs.~\cite{FileviezPerez:2010gw,Dulaney:2010dj,Chao:2010mp,Duerr:2013dza} for anomaly cancellations in detail. 
 
\begin{table}[t]
\centering
\begin{tabular}{c|c|c|c|c|c|c|c|c|c|c|c|c|c|c}
\hline \hline scenario &symmetries  & $Q_L$ &$\ell_L$ & $U_R$ & $D_R$ &$E_R$& $N_R$ & $H $ & $\psi_L$ &$\psi_R$ & $\chi_L$ & $\chi_R$&$\eta_L$ &$\eta_R$\\
\hline
(i) & $U(1)_{\mathbf{B-L}}$ &  $+\frac{1}{3}$ &$-1$ &$+\frac{1}{3}$ &$+\frac{1}{3}$&$-1$ &$-1$ & $0$ &  $\times$ & $\times$  & $\times$ & $\times$ & $\times$ & $\times$  \\ 
\hline
(ii) & $U(1)_{\mathbf{R}}$ &  $0$ &0 & $-1$ & $+1$ & $-1$ &$+1$ & 1 &  $\times$ & $\times$  & $\times$ & $\times$ & $\times$ & $\times$  \\ 
\hline
(iii) & $U(1)_{\mathbf{B}}$  &  $+\frac{1}{3}$ & 0 &$ +\frac{1}{3} $ &$ +\frac{1}{3}$&0 &0 & 0 &   $-1$ & $+2$ & $+2$ & $-1$ & $+2$ & $-1$ \\ 
\hline
(iv) & $U(1)_{\mathbf{L}}$  &  $0$ &$+1$ &$0$ &$0$&$+1$ & $+1$ & 0 &  $-1$  & $+2$ & $+2$ & $-1$ & $+2$ & $-1$  \\ 
\hline
\hline
\end{tabular}
\caption{ Quantum numbers of  various fields under the $U(1)_{\mathbf{B-L}}^{},~U(1)_{\mathbf{R}},~U(1)_{\mathbf{B}} $ and $U(1)_{\mathbf{L}}$,  where $Q_L$, $\ell_L$, $U_R$, $D_R$, $E_R$, $N_R$ correspond to left-handed quark doublets, left-handed lepton doublets, right-handed up-type quarks, right-handed down-type quarks, right-handed charged leptons, and right-handed neutrinos, respectively, $H$ is the SM Higgs doublet, $\psi_{L,R}$ is new vector-like lepton doublets, $\xi_{L,R}$ is vector-like charged fermions, $\eta_{L,R}$ is vector-like neutral fermions, required for anomaly cancellations.  ``$\times$" means the no need of corresponding particle. }\label{table1}
\end{table}

Now we consider the triangle anomalies for  chiral fermions in various U(1) models. We follow the definition of the baryon current and the lepton current as  $j_B^\mu = {1\over 3} \sum_f \left( \bar Q \gamma^\mu Q + \bar u_R \gamma^\mu u_R^{}\right.$ + $\left.\bar d_R^{} \gamma^\mu d_R^{}  \right)$,  $J_L^\mu = \sum_f \left( \bar \ell_L \gamma^\mu \ell_L^{} + \bar E_R^{} \gamma^\mu E_R^{} \right)$ and $J_N^\mu=\sum_f\bar N_R^{} \gamma^\mu N_R^{} $. For the $U(1)_{\mathbf{B-L}}$, the triangle anomalies  for each flavor of left-handed fermion doublets and right-handed fermion singlets can be written as 
\bea
\partial_\mu \left( j^\mu_{B, Q} \right)&=& \frac{1}{32 \pi^2 } \left(  g^2 W \widetilde W + \frac{1}{9}g^{\prime 2} F \widetilde{F} +\frac{4}{9} g^{ 2 }_{B-L} F^\prime \widetilde{F}'\right)\\
\partial_\mu \left( j^\mu_{B, u} \right)&=& \frac{1}{16 \pi^2 } \left(  - \frac{4}{9}g^{\prime 2} F \widetilde{F} -\frac{1}{9} g^{2 }_{B-L} F^\prime \widetilde{F}'\right) \\
\partial_\mu \left( j^\mu_{B,d} \right)&=& \frac{1}{16 \pi^2 } \left(   -\frac{1}{9}g^{\prime 2} F \widetilde{F} -\frac{1}{9} g^{ 2 }_{B-L} F^\prime \widetilde{F}'\right)\\
\partial_\mu \left( j^\mu_{L, \ell} \right)&=& \frac{1}{32 \pi^2 } \left(  g^2 W \widetilde W +g^{\prime 2} F \widetilde{F} +4  g^{ 2 }_{B-L} F^\prime \widetilde{F}'\right)\\
\partial_\mu \left( j^\mu_{L, E} \right)&=& \frac{1}{16 \pi^2 } \left( - g^{\prime 2} F \widetilde{F} - g^{ 2 }_{B-L} F^\prime \widetilde{F}'\right) \\
\partial_\mu \left( j^\mu_{L,N} \right)&=& \frac{1}{16 \pi^2 } \left( -g^{ 2 }_{B-L} F^\prime \widetilde{F}'\right)
\eea
where $W\widetilde{W}$, $F\widetilde{F}$ and $F^\prime \widetilde{F}^\prime$ are Chern-Simons term for the $SU(2)_L$ gauge field,  the $U(1)_Y$ gauge field and the the $U(1)_{\mathbf{B-L}}$ gauge field, respectively.  For  the $U(1)_{\mathbf{R}}$ case, triangle anomalies for left-handed fermions are the same as these in the SM, as left-handed fermions carry null $U(1)_{\mathbf{R}}$ charge. Triangle anomalies for right-handed fermions take the following form
\bea
\partial_\mu \left( j^\mu_{B, u} \right)&=& \frac{1}{16 \pi^2 } \left(  - \frac{4}{9}g^{\prime 2} F \widetilde{F} -g^{2 }_{R} F^\prime \widetilde{F}'\right) \\
\partial_\mu \left( j^\mu_{B,d} \right)&=& \frac{1}{16 \pi^2 } \left(   -\frac{1}{9}g^{\prime 2} F \widetilde{F} - g^{ 2 }_{R} F^\prime \widetilde{F}'\right)\\
\partial_\mu \left( j^\mu_{L, E} \right)&=& \frac{1}{16 \pi^2 } \left( - g^{\prime 2} F \widetilde{F} - g^{ 2 }_{R} F^\prime \widetilde{F}'\right) \\
\partial_\mu \left( j^\mu_{L,N} \right)&=& \frac{1}{16 \pi^2 } \left( -g^{ 2 }_{R} F^\prime \widetilde{F}'\right)
\eea
For the $U(1)_{\mathbf{B}}$, triangle anomalies for quarks are the same as these in the Eqs. (2-5) up to the replacement of gauge couplings, $g^{}_{\mathbf{B-L}} \to g_B^{}$, while triangle anomalies for leptons are the same as these  in the SM. In the meanwhile, triangle anomalies for new chiral fermions takes the following form 
\bea
\partial_\mu^{} \left(  j^\mu_{\psi_L}\right) &=& + {1\over 32\pi^2 } \left( g^2 W \widetilde{W} + g^{\prime 2} F \widetilde{F} + 4 g_B^2 F^\prime \widetilde{F}^\prime\right) \\
\partial_\mu^{} \left(  j^\mu_{\psi_R}\right) &=&- {1\over 32\pi^2 } \left( g^2 W \widetilde{W} + g^{\prime 2} F \widetilde{F} + 16 g_B^2 F^\prime \widetilde{F}^\prime\right) \\
\partial_\mu^{} \left(  j^\mu_{\chi_L}\right) &=&+ {1\over 16\pi^2 } \left(  g^{\prime 2} F \widetilde{F} + 4 g_B^2 F^\prime \widetilde{F}^\prime\right) \\
\partial_\mu^{} \left(  j^\mu_{\chi_R}\right) &=&- {1\over 16\pi^2 } \left(  g^{\prime 2} F \widetilde{F} +  g_B^2 F^\prime \widetilde{F}^\prime\right) \\
\partial_\mu^{} \left(  j^\mu_{\eta_L}\right) &=&+ {1\over 4\pi^2 } \left(   g_B^2 F^\prime \widetilde{F}^\prime\right) \\
\partial_\mu^{} \left(  j^\mu_{\eta_R}\right) &=&- {1\over 16\pi^2 } \left(  g_B^2 F^\prime \widetilde{F}^\prime\right) 
\eea
from which, one has
\bea
\partial_\mu \left(  j^\mu_{\psi_L}+j^\mu_{\psi_R}+j^\mu_{\chi_L}+j^\mu_{\chi_R}+j^\mu_{\eta_L}+j^\mu_{\eta_R} \right) =0 \; .
\eea
For the $U(1)_{\mathbf{L}}$,  triangle anomalies for leptons are the same as these in Eqs. (5-7), triangle anomalies for quarks are the same as these in the SM, and triangle anomalies for new fermions are the same as these in Eqs. (13-17).  

From above current equations, one can conclude that 
\bea
\partial_\mu \left(j^\mu_B -j^\mu_L  \right) =  \partial_\mu j^\mu_N
\eea
for these  U(1) symmetric extension models, which means that triangle anomalies for right-handed neutrinos  can be a source of the baryon asymmetry in the case where right-handed neutrinos never reach thermal equilibrium in the early universe.

\section{gauge field production in inflation via Chern-Simons term}
In this section we review the mechanism of the gauge field production in the model of axion-inflation~\cite{Turner:1987bw,Garretson:1992vt,Anber:2006xt}.  For simplicity, we consider the cosmic inflation driven by axion-like pseudo-scalar, $\phi$, that couples to a new gauge field $A^\prime_\mu$ via an effective Chern-Simons term. The relevant action is 
\bea
S= \int d^4 x \left\{ \sqrt{-g} \left[ \frac{1}{2} g^{\mu\nu}_{} \partial_\mu \phi \partial_\nu \phi -V(\phi)\right] -{\alpha \over 4\pi} {\phi \over f_a} F^\prime_{\mu\nu} \widetilde{F}^{\prime\mu\nu}\right\}
\eea
where $\alpha$ is the structure constant of the gauge field, i.e.,  $\alpha=g^{ 2}_X /(4\pi)$ with $g_X^{}$ the gauge coupling of the $U(1)_X$ symmetry, $F^\prime_{\mu\nu}$ denotes the field strength tensor,  $\widetilde{F}^{\prime\mu\nu} =\varepsilon^{\mu\nu\rho\sigma} F_{\rho \sigma}^{\prime}/2$ with $\varepsilon^{0123}=1$. We take the Friedmann-Robertson-Walker (FLRW) metric, $ds^2 =dt^2 -a^2(t) dx^2$ where $a$ is the scale factor, and use the conformal time $d\eta = dt /a$ in the following analysis for simplicity.

Taking the variation with respect to the new gauge field $A^\prime$, one gets its equation of motion in the form of the modified Bianchi Identity. The quantization of the gauge field can be performed in the conventional way with the plane wave expansion of the $\bm A$ given by
\bea
A(t, x) =\sum_{\lambda=\pm} \int {d^3 \bm k \over (2\pi)^3 } \left[ A_\lambda (\eta, k) \bm \varepsilon_\lambda (\bm k ) a_\lambda^{} (\bm k) e^{i \bm k \cdot \bm x } + {\rm h.c.}\right]
\eea
where $\varepsilon_\lambda (\bm k)$ is the polarization vector fulfilling the following relations: $\varepsilon_\lambda^* (\bm k) \cdot \varepsilon^{\lambda^\prime}_{}(\bm k) =\delta_{\lambda \lambda^\prime}$, $\varepsilon_\lambda (\bm k) \cdot {\bm k} =0$ and $i {\bm k} \times \varepsilon_\lambda (\bm k) = \lambda |\bm k| \varepsilon_\lambda (\bm k)$,  $a_\lambda^{} (\bm k)$ and $a^\dagger (\bm k)$ are the annihilation and creation operators satisfying the canonical commutation relation, $[a_\lambda^{} (\bm k), a^\dagger_{\lambda^\prime} (\bm k^\prime)] =(2\pi)^3 \delta_{\lambda \lambda\prime}^{} \delta^{(3)}(\bm k -\bm k^\prime)$, $A_\lambda$ denotes the mode function. As a result, the equation of motion for the gauge field in Fourier space takes the following form
\bea
\left[ \frac{\partial^2 }{\partial \eta^2} + k(k +2\lambda \xi  aH )\right] A_{\lambda} (\eta, k) =0
\eea
where $\lambda=\pm$, $H=\dot a/a$ being the Hubble parameter, $\eta$ is the conformal time, and 
\bea
\xi \equiv { \alpha \lambda \dot \phi \over 2\pi f_a H} \; ,
\eea
which is related to the slow-roll parameter $\varepsilon$, via $\xi \sim \lambda M_{pl} \sqrt{\varepsilon}/ (\sqrt{2} fH)$~\cite{Maleknejad:2020yys,Maleknejad:2020pec}.
The mode equation is isotropic in momentum space.   To solve the equation~(22), we require the mode $A_\lambda$ reduces to the Bunch-Davis solution~\cite{Weinberg:2008zzc} in the asymptotic past, i.e., $A_\lambda (k,\eta) = {e^{-i k\eta} \over \sqrt{2k}}$ for $-k\eta \to \infty$. In the limit of the constant $\xi$, which is consistent with the slow-roll approximation,  the analytical solution is given by~\cite{Maleknejad:2016qjz}  
\bea
A_\lambda(\eta, k)= \frac{e^{ \lambda \pi \xi /2}}{\sqrt{2k}} W_{-i\lambda \xi, 1/2}^{} (2ik \eta )
\eea
where $W_{-i\lambda \xi, 1/2}$ is the Whittaker function. It may lead to the generation of the Chern-Simons number~\cite{Domcke:2018eki}, and may also lead to generation of chiral fermions via the back-reaction~\cite{Maleknejad:2018nxz}.  We refer the reader to Refs.~\cite{Maleknejad:2018nxz,Domcke:2018eki,Adshead:2018oaa,Domcke:2019mnd} and references cited therein,  for the calculation of gauge fields and fermions production during the axion inflation in detail. 


\section{ {\it Baryogenesis} }

Plugging the mode $A_\lambda (k, \eta)$ into $F^\prime\widetilde{F^\prime}$, one may estimate the change of the Chern-Simons number as well as the number density of various chiral fermions from  their current equations.  We assume that the new U(1) gauge symmetry is spontaneously broken shortly after the inflation and the helical gauge field is totally transported into the SM fermions  during the reheating time, which is similar to the assumption in the  gravitational wave Leptogenesis mechanism~\cite{Alexander:2004us}. Actually, a specific helical gauge field may survive until to the scale of the  U(1) gauge symmetry spontaneous breaking, which is the case of the Hyer-magnetic field as shown in the Refs.~\cite{Kamada:2016cnb,Jimenez:2017cdr,Domcke:2019mnd}.  To estimate the Chern-Simons number, one may first reform the term $F^\prime \widetilde F^\prime$ into a total derivative, $2\partial_\mu K^\mu$,  and take $K^0$ as the Chern-Simons number~\cite{Maleknejad:2020yys,Maleknejad:2020pec},
\bea
n_{CS} \equiv \frac{1}{(2\pi)^2} \mathcal{K} (\xi) a^3 H^3  =\frac{1}{(2\pi)^2} \sum_{\lambda=\pm} \lambda e^{i\kappa_{\lambda}\pi} \int \tilde{\tau}^3 d\ln\tilde{\tau} W^{*}_{\kappa_{\lambda},\mu}(-2i\tilde{\tau})W_{\lambda_{\sigma},\mu}(-2i\tilde{\tau}) a^3 H^3
\eea 
in which one has defined the function $\mathcal{K} (\xi)$.  

Further taking  the current equation for various chiral fermions as 
$
\partial _\mu \left(\sqrt{-g} J_{f, \sigma} ^\mu \right) = \epsilon_\sigma N_{f, \sigma}  \frac{g_X ^2}{16\pi ^2} F^\prime_{\mu \nu} \widetilde{F}^{\prime\mu \nu}_{ }
$, where $f$ indicates the flavor of the fermion, $\sigma=L,R$ indicating the chirality of the fermion, $\epsilon_L =+ $  and  $\epsilon_R =- $,  $N_{f, \sigma}$ can be read from eqs.~(2-17), then the number density of a chiral fermion can be given as
\bea
n_{f, \sigma} = -\epsilon_\sigma N_{f,\sigma} \frac{g_X ^2}{8 \pi^2 a^3} n_{CS} = -\epsilon_i N_i \frac{g_X ^2}{2(2\pi)^4}H^3 \mathcal{K}(\xi)
\eea 
which provides the initial inputs for the transport equations of the SM particles. 

To estimate the final baryon asymmetry, we replace the number density $n_i$ of various particles with their chemical potential $\mu_i$: $n_i (t) =g_i \mu_i(t) T^2 /6$, where $T$ is the temperature of the plasma, $g_i$ represents the degrees of freedom. Transport equations in the radiation dominate epoch can be written as
\bea
- \frac{d}{d \ln T} \left(\frac{\mu _i}{T}\right) = - \frac{1}{g_i} \sum_\alpha n^\alpha _i \frac{\gamma_\alpha}{H} \left( \sum_j n^\alpha_j \frac{\mu_j}{T} \right)
\eea
where $ n^\alpha_i$ is the charge of the species $i$ in the operator $O_\alpha$, $\gamma^\alpha$ is the interaction rate per unit time for operator $O_\alpha$. We will only consider effects of the  weak sphaleron, the strong sphaleron and the Yukawa interactions in our analysis. The full set of the transport equations are listed in the appendix A. We assume neutrinos are Dirac particles and the tiny but non-zero neutrino masses are generated by the conventional Yukawa interaction  in our analysis, which means that the Yukawa interaction of right-handed neutrinos never reach equilibrium in the early universe.

After the evolution of the transport equations from the reheating temperature  to the electroweak weak scale, after which the electroweak sphaleron is quenched,  one derives the chemical potential for various quarks which can be translated into the baryon number density $n_B$. The final BAU is then
\bea
\eta_B =\frac{n_B}{n_\gamma} =\frac{28}{97} \frac{n_{N-L}}{n_\gamma} \; ,
\eea
where $n_\gamma =\frac{2 \zeta(3)}{\pi^2} T^2$, being the number density of the photon, with the Riemann zeta function $\zeta(3) \approx 1.202$. In the following, we will study the BAU generated in various U(1) models,  separately. 

\begin{figure}[t]
\includegraphics[width=0.45\textwidth]{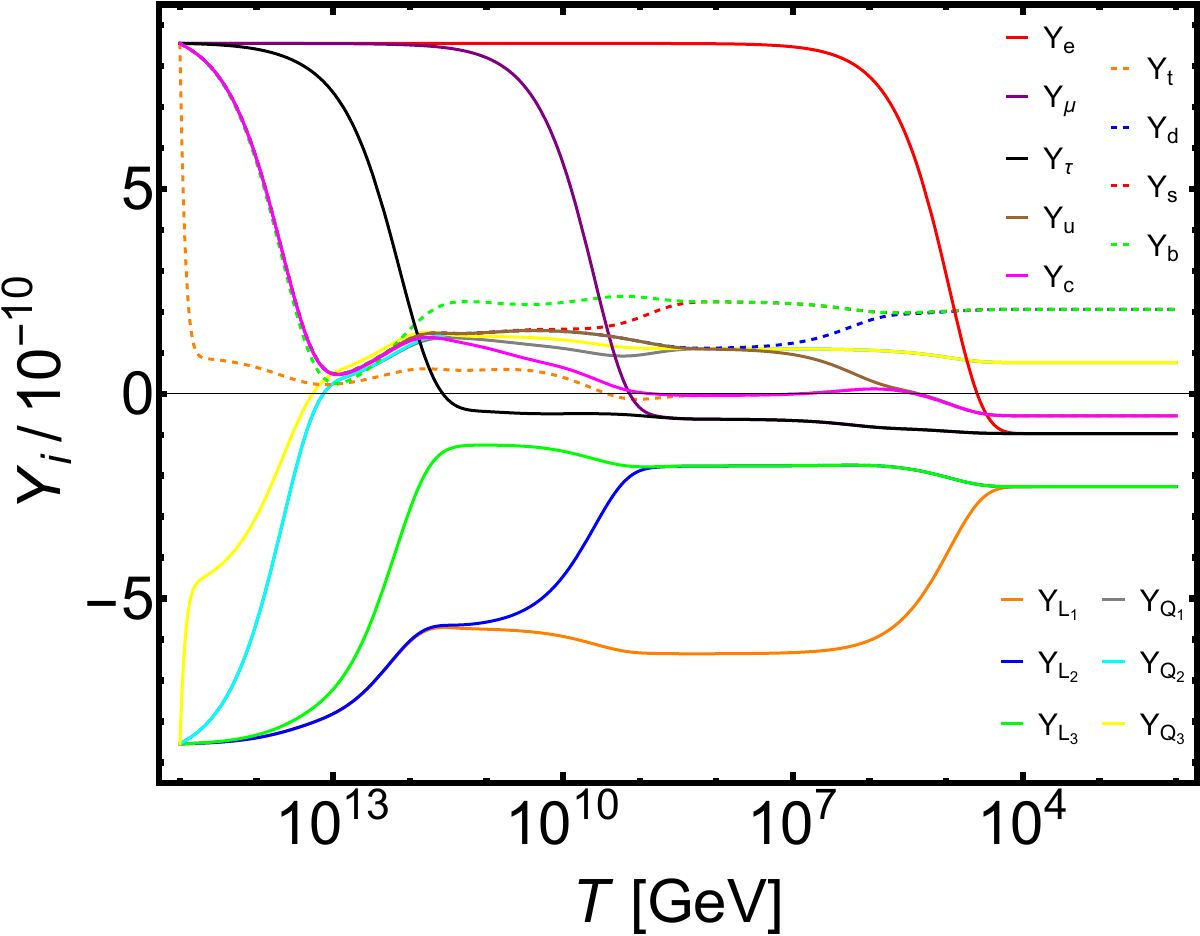}
\hspace{0.5cm}
\includegraphics[width=0.465\textwidth]{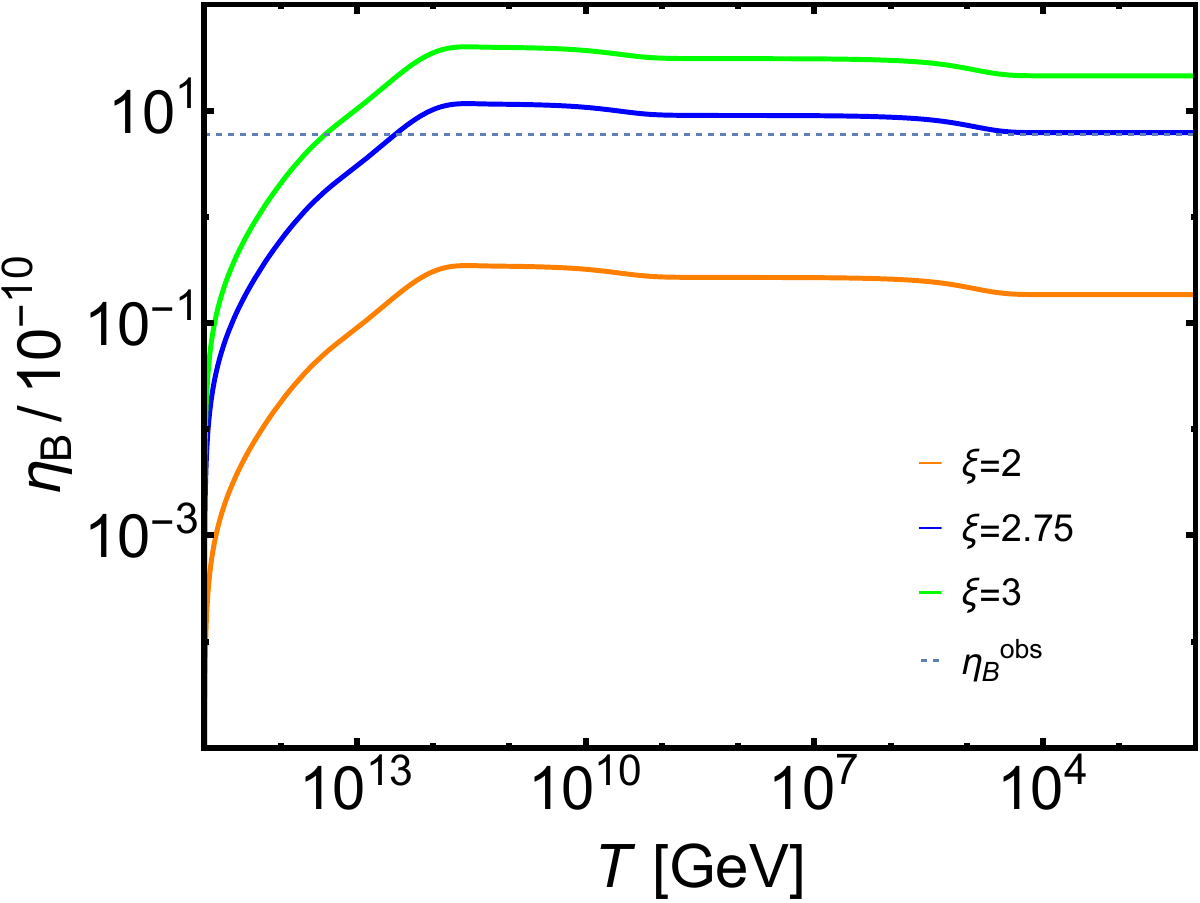}
\caption{ Left-panel: various chemicals divided by the temperature as the function of the temperature, where the orange, blue, green, red, purple, black, gray, cyan, yellow, brown, magenta, dashed-orange, dashed-blue, dashed-red, dashed-green curves corresponds the running behavior of three generations of left-handed lepton doublets ($L_1, ~L_2, ~L3$), right-handed lepton singlets, left-handed quark doublets ($Q_1,~Q_2, ~Q_3$) and right-handed up, charm, top, down ,strange, bottom quark, respectively.  Right-panel: the  BAU generated by the gauge field of the $U(1)_{\mathbf{B-L}}$ with the solid, dashed and dotted curves correspond to the inflation parameter $\xi=2, 2.75, 3$ respectively.  The horizontal dashed line corresponds to the observed BAU.} \label{figbml}
\end{figure}

\framebox{\large\bf $U(1)_{\mathbf{B-L}}$:} For the $U(1)_{\mathbf{B-L}}$ case, both asymmetric baryon number and asymmetric lepton number are generated by the $F^\prime \widetilde{F}^\prime $ term during the reheating time, but the total $B-L_{\slashed{N}} -L_N$ is conserved at $T_i$, which is the temperature of the reheating.  Remembering that right-handed  Dirac neutrinos do not enter thermal equilibrium,  nonzero $B-L_{\slashed{N}} $ and $L_N$ evolve separately in the early universe,  resulting in non-zero BAU at the electroweak scale.  We show in the left panel of the Fig. \ref{figbml}  chemicals of various species divided by the temperature, namely $Y_i =\mu_i/T$,  as the function of the temperature. For the physical meaning of each curve, see caption of the Fig.~\ref{figbml} for detail. Each Inflection point  in the figure corresponds to  a specific interaction entering into the thermal equilibrium.   Obviously, the shape of a curve is consistent with the analytical estimation of  the temperature at which an interaction enters thermal equilibrium. We show in the right-panel of the Fig.~\ref{figbml} the BAU generated by the gauge field of the $U(1)_{\mathbf{B-L}}$ as the function of the temperature, with the solid, dashed and dotted curves corresponding to $T_i =10^{15}~{\rm GeV}$ and $\xi=2,2.75,3$, respectively. The horizontal blue solid line corresponds to the observed BAU. Obviously, the observed BAU can be addressed in this scenario by setting an appropriate  inflation parameter $\xi$ and a reheating temperature. 

\begin{figure}[t]
\includegraphics[width=0.45\textwidth]{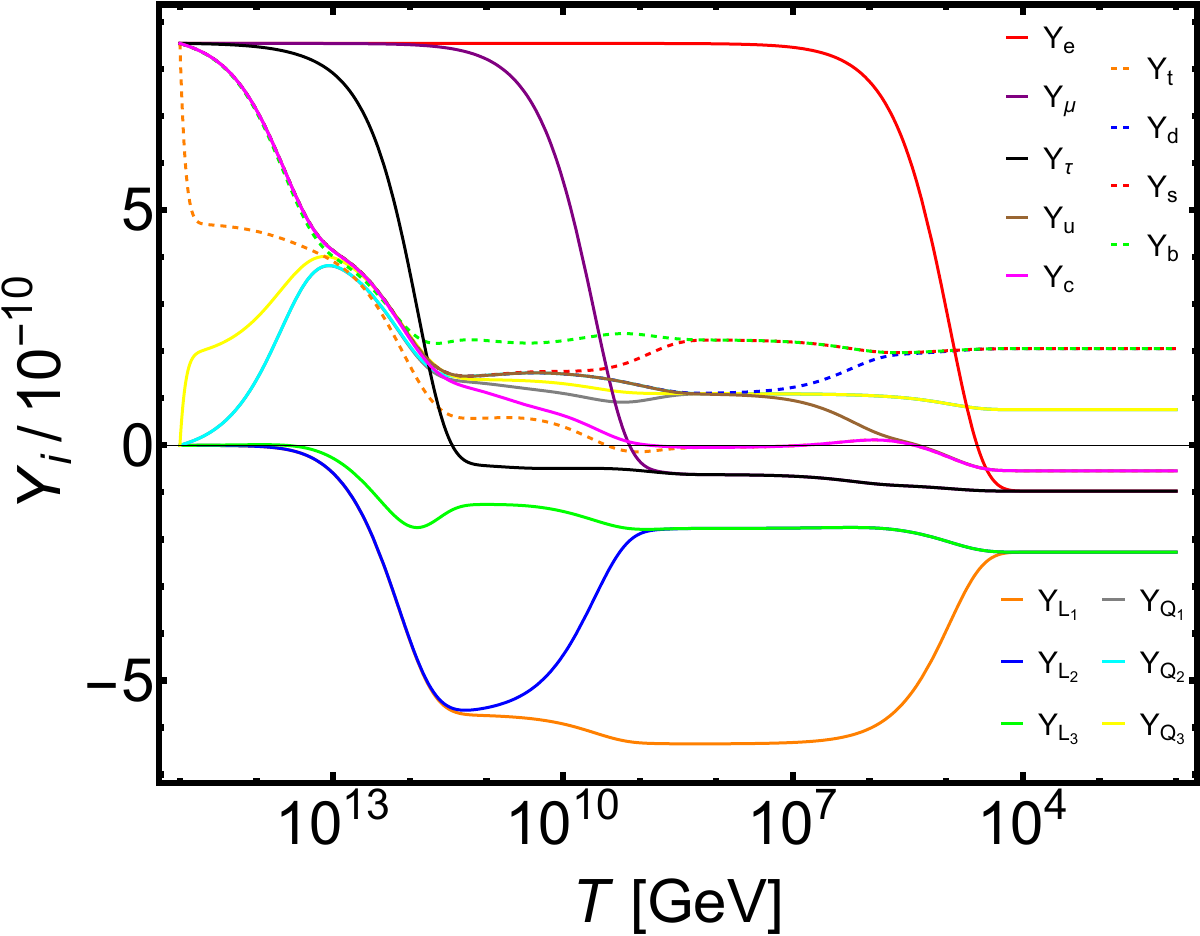}
\hspace{0.5cm}
\includegraphics[width=0.46\textwidth]{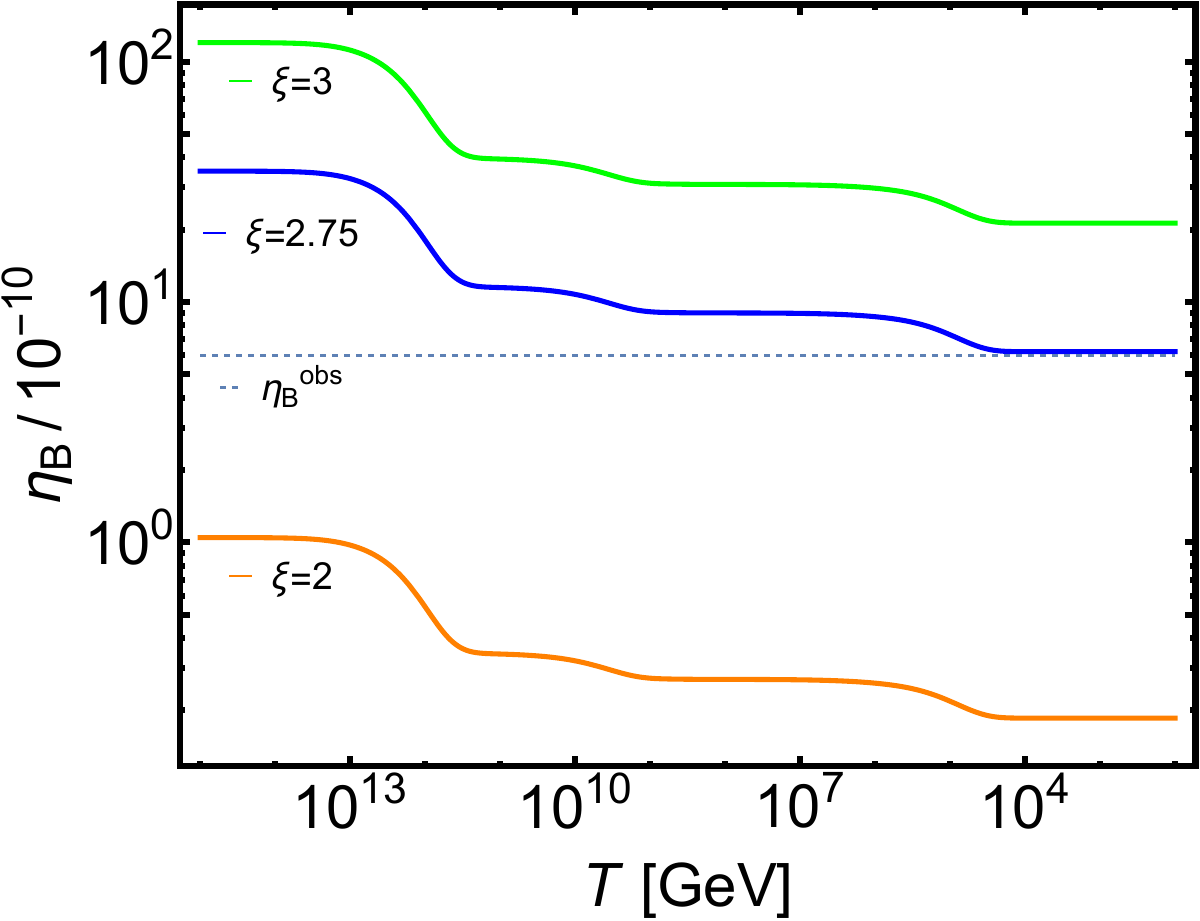}
\caption{Left-panel: $Y_i$ as the function of the temperature, with physical meaning of each curve the same as these in the Fig.~\ref{figbml}.  Right-panel: the BAU  as the function of the temperature in the $U(1)_R$. }\label{figu1R}
\end{figure}

\framebox{\large\bf $U(1)_{\mathbf{R}}$:} For the $U(1)_R$ case, initial asymmetries for both  right-handed charged leptons  and right-handed quarks are generated during the reheating. These asymmetries for right-handed fermions are diffused into asymmetries of left-handed fermions via Yukawa processes as well as the strong sphaleron process, which is subsequently transported into the BAU via the electroweak sphaleron process.  
We show in the left-panel of the Fig.~\ref{figu1R}, $Y_i$ as the function of the temperature for various species, in which right-handed fermions have universal initial density, while the initial density for left-handed fermions are zero. The shape of each curve is governed by the temperature at which the corresponding interaction enters thermal equilibrium. The  curves of the up-quark and the down-quark number densities overlap with each other, which is because their Yukawa interactions enter thermal equilibrium at almost the same temperature.
We show in the right-panel of the Fig. 2 the BAU  as the function of the temperature in the $U(1)_R$ case, using the same initial inputs  as these in the Fig. 1.   To numerically fit the observed BAU, the inflation parameter should be set to be  $\xi=2.75$ and $T_i =10^{14}~{\rm GeV}$ in this case. 

\begin{figure}[t]
\includegraphics[width=0.45\textwidth]{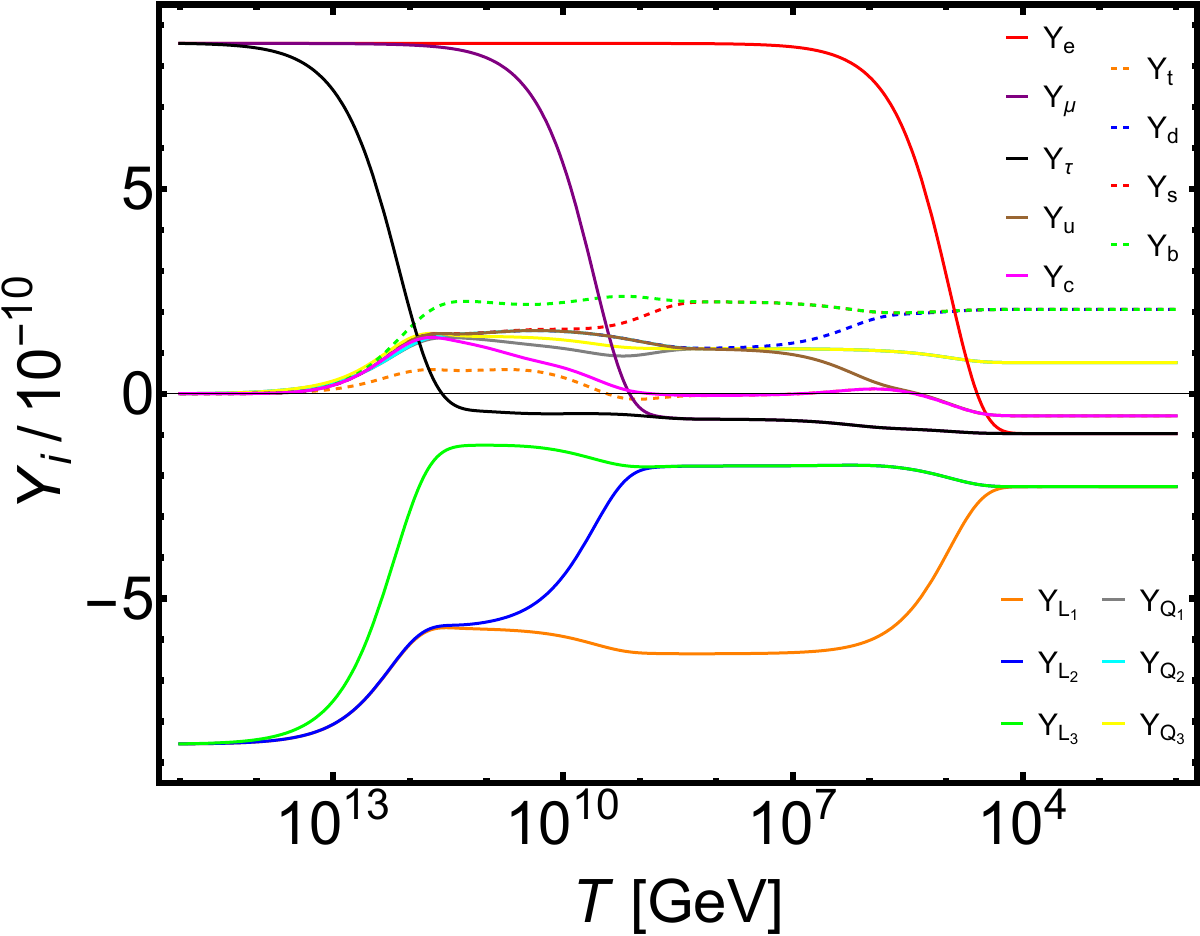}
\hspace{0.5cm}
\includegraphics[width=0.46\textwidth]{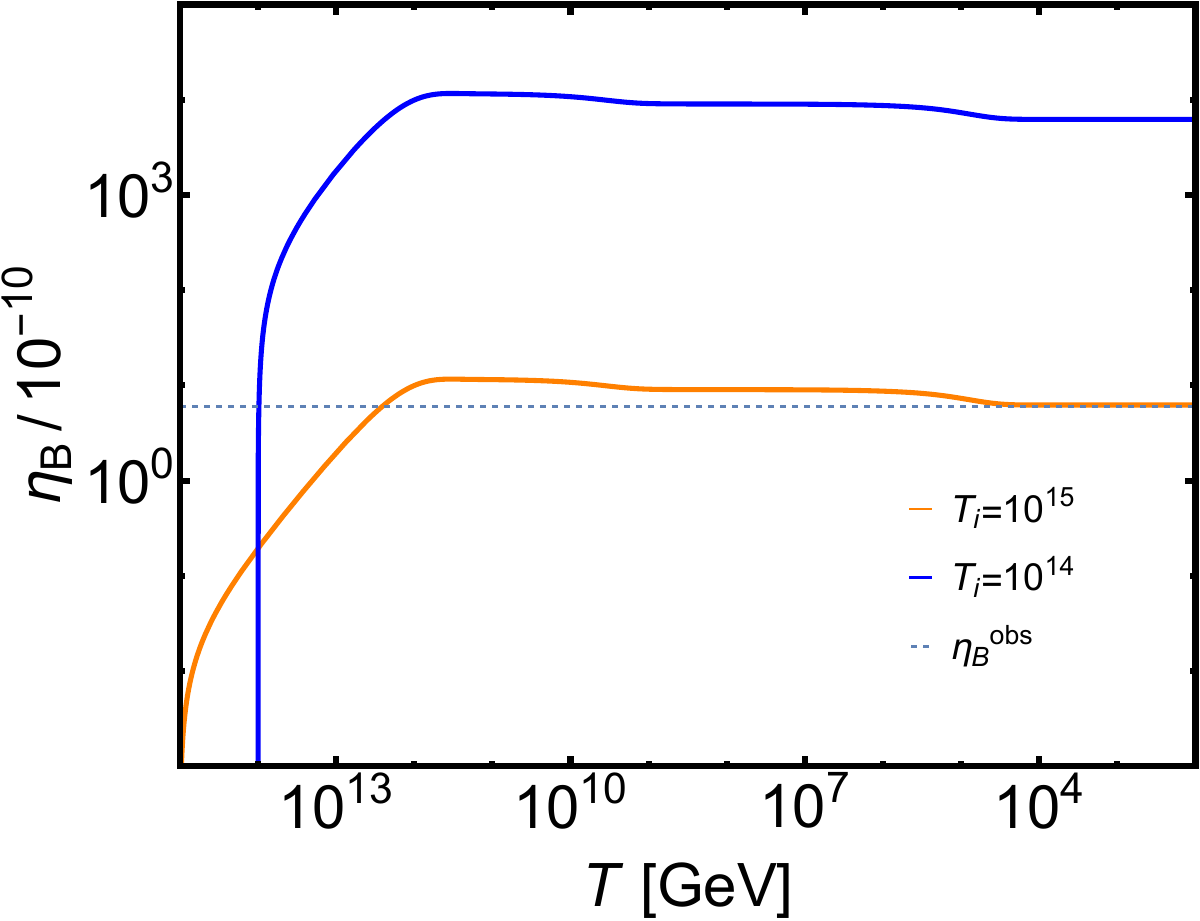}			
\caption{Left-panel: $Y_i$ as the function of the temperature, with physical meaning of each curve the same as these in the Fig.~\ref{figbml}.  Right-panel: the BAU  as the function of the temperature in the $U(1)_L$ with the orange and blue curves correspond to $T_i=10^{14}$ GeV and $10^{15}$ GeV respectively. } \label{figL}
\end{figure}

\framebox{\large\bf $U(1)_{\mathbf{B}}$:} For the $U(1)_{\mathbf{B}}$ case, initial asymmetries for left-handed quark doublets and right-handed quark singlets can be generated, but the total baryon asymmetry is zero at $T_i$. Considering that  no lepton asymmetry is generated from the axion inflation, none BAU can be generated at the electroweak sale.  However, it should be mentioned that this scenario might work in the existence of the type-I seesaw model and the BAU might be generated via the wash-in Leptogenesis mechanism~\cite{Domcke:2020quw}, in which the electroweak sphaleron process partially transfers the quark asymmetry into the lepton asymmetry, and then the lepton asymmetry is washed-out by the lepton-number-violating interactions mediated by right-handed Majorana neutrinos, resulting in non-zero B-L asymmetry at a temperature when right-handed neutrinos are decoupled and some of right-handed quarks does not enter into the thermal equilibrium.

\framebox{\large\bf $U(1)_{\mathbf{L}}$:} For the $U(1)_L$ case, both nonzero $L_{\slashed{N}}$ and $L_{N}$ can be generated, and the initial baryon number is zero. $L_{\slashed{N}}$ can be transported into asymmetries of left-handed lepton doublets, which is then transported into the baryon asymmetry via the weak sphaleron process. $L_{N}$  is kept to now-days resulting in a non-zero neutrino asymmetry, which is the same as cases of  $U(1)_{\mathbf{B-L}}$ and $U(1)_{\mathbf R}$.   
The evolution behaviors of various species are shown in the left-panel of the Fig.~\ref{figL}, with physical meaning of each curve the same as these in the Fig.~\ref{figbml}. 
We show in the right-panel of the Fig.~\ref{figL} the BAU as the function of the temperature by setting the inflation parameter $ \xi=2.75$, $T_i=10^{14}$ GeV (blue) and $10^{15}$ GeV (orange),  respectively. 
It can be read from the plot that the observed BAU can also be generated in this case with proper inputs of inflation parameters. 

\section{Concluding remarks}

The BAU  is a longstanding question in the particle physics and cosmology. In this paper we revisited the axion-inflation baryogenesis mechanism by extending the SM with various U(1) gauge symmetries and the assuming that new gauge field couples to the axion-inflaton via the Chern-Simons coupling, which provides the source of C and CP violations. We find that a net B-L asymmetry can be generated during the reheating temperature for the $U(1)_{\mathbf{B-L}}$, $U(1)_L$ and $U(1)_R$ cases, whenever neutrinos are Dirac particles and its Yukawa interactions never reach thermal equilibrium in the early universe, which is similar to the case of the Dirac Leptogenesis.  The net B-L asymmetry can then be transported into the BAU via the electroweak sphaleron process. Our numerical results show that the observed BAU can be addressed in those scenarios by proper inputs of inflation parameters.  This study provides a workable baryogenesis mechanism and the U(1) model used in this study might be tested in the future high energy collider experiments as well as  in astronomical observations.  It should be mentioned that one may also investigate the BAU generated by the wash-in Leptogenesis mechanism in our axion-inflation models, which,  although interesting but beyond the reach of this paper,  will be presented in a future study.				
\begin{acknowledgments}
This work was supported in part by the National Key R\&D Program of China under Grant No. 2023YFA1607104, by the National Natural Science Foundation of China under Grants No. 11775025 and No. 12175027, and by the Fundamental Research Funds for the Central Universities under Grant No. 2017NT17.
\end{acknowledgments}

\appendix 
\section{Transport equations}
The full set of transport equations can be find in various references. We list them here for the references:
\bea
-\frac{d}{d\ln T}\left( \frac{\mu_{E_k}}{T}\right)&=&+\frac{1}{g_{E_k}}\frac{\gamma_{Y_{E_k}}}{H}\left(-\frac{\mu_{E_k}}{T}+\frac{\mu_{L_k}}{T}-\frac{\mu_H}{T}\right) ,\\
-\frac{d}{d\ln T}\left( \frac{\mu_{L_k}}{T}\right)&=&-\frac{1}{g_{L_k}}\frac{2\gamma_{WS}}{H}\left[\sum_i \frac{\mu_{L_{i}}}{T}+3 \sum_i \frac{\mu_{Q_{i}}}{T} \right]\nonumber\\
&&-\frac{1}{g_{E_k}}\frac{\gamma_{Y_{E_k}}}{H}\left(-\frac{\mu_{E_i}}{T}+\frac{\mu_{L_i}}{T}+\frac{\mu_H}{T}\right),\\
-\frac{d}{d\ln T}\left( \frac{\mu_{U_{k }}}{T}\right)&=&\frac{1}{g_{U_{k}}}\frac{4\gamma_{SS}}{H}\left[-\sum_{j=1}^3 \left( \frac{\mu_{U_{j}}}{T}+\frac{\mu_{D_{j}}}{T} \right)+\sum_{i=1}^3\frac{2\mu_{Q_{i}}}{T} \right]\nonumber\\
&& +\frac{1}{g_{U_k}}\frac{\gamma_{Y_{U_k}}}{H}\left[-\frac{\mu_{U_k}}{T}+\frac{\mu_{Q_k}}{T}+\frac{\mu_H}{T} \right] \; , \\
-\frac{d}{d\ln T}\left( \frac{\mu_{D_{k }}}{T}\right)&=&+\frac{1}{g_{D_{k}}}\frac{\gamma_{SS}}{H}\left[-\sum_{j=1}^3 \left( \frac{\mu_{U_{j}}}{T}+\frac{\mu_{D_{j}}}{T} \right)+\sum_{i=1}^3\frac{2\mu_{Q_{i}}}{T} \right]\nonumber\\
&& +\frac{1}{g_{D_k}}\frac{\gamma_{Y_{D_k}}}{H}\left[-\frac{\mu_{D_k}}{T}+\frac{\mu_{Q_k}}{T}-\frac{\mu_H}{T} \right] \\
-\frac{d}{d\ln T}\left( \frac{\mu_{Q_k}}{T}\right)&=&-\frac{1}{g_{Q_k}}\frac{2\gamma_{SS}}{H}\left[-\sum_{j=1}^3 \left( \frac{\mu_{U_{j}}}{T}+\frac{\mu_{D_{j}}}{T} \right)+\sum_{i=1}^3\frac{2\mu_{Q_{i}}}{T}\right]
\nonumber\\
&&-\frac{1}{g_{Q_k}}\frac{3\gamma_{WS}}{H}\left[\sum_i \frac{\mu_{L_{i}}}{T}+3 \sum_i \frac{\mu_{Q_{i}}}{T}\right]-\frac{1}{g_{Q_k}}\frac{\gamma_{Y_{U_k}}}{H}\left[-\frac{\mu_{U_k}}{T}+\frac{\mu_{Q_k}}{T}+\frac{\mu_{H}}{T}\right]\nonumber\\
&&-\frac{1}{g_{Q_k}}\frac{\gamma_{Y_{D_k}}}{H}\left[-\frac{\mu_{D_k}}{T}+\frac{\mu_{Q_k}}{T}-\frac{\mu_{H}}{T}\right] \\
-\frac{d}{d\ln T}\left( \frac{\mu_H}{T}\right)&=&-\frac{1}{g_H} \sum_k \left( \frac{\gamma_{Y_{E_k}}}{H}\left[-\frac{\mu_{E_k}}{T}+\frac{\mu_{L_k}}{T}+\frac{\mu_H}{T} \right]- \frac{\gamma_{Y_{D_k} }}{H}\left[-\frac{\mu_{D_k} }{T}+\frac{\mu_{Q_k}}{T}-\frac{\mu_{H}}{T}\right] \right)
\nonumber\\
&&-\frac{1}{g_H}\sum_k \frac{\gamma_{Y_{U_k}}}{H}\left[-\frac{\mu_{U_k}}{T}
+\frac{\mu_{Q_k}}{T}+\frac{\mu_{H}}{T} \right]
\eea
where $E_k$, $U_k$, $D_k$, $L_k$ and $Q_k$, with k=1,2,3,  represent right-handed charged leptons, right-handed up-type quarks, right-handed down-type quarks, left-handed lepton doublets and left-handed quark doublets, respectively,  the transport rate per unit time $\gamma_{\alpha}\equiv{6\Gamma_{\alpha}}/{T^{3}}$ are given in the Table 1 of Ref.~\cite{Domcke:2020kcp}.

\bibliography{AinflationBG}

\begin{thebibliography}{70}%
\makeatletter
\providecommand \@ifxundefined [1]{%
 \@ifx{#1\undefined}
}%
\providecommand \@ifnum [1]{%
 \ifnum #1\expandafter \@firstoftwo
 \else \expandafter \@secondoftwo
 \fi
}%
\providecommand \@ifx [1]{%
 \ifx #1\expandafter \@firstoftwo
 \else \expandafter \@secondoftwo
 \fi
}%
\providecommand \natexlab [1]{#1}%
\providecommand \enquote  [1]{``#1''}%
\providecommand \bibnamefont  [1]{#1}%
\providecommand \bibfnamefont [1]{#1}%
\providecommand \citenamefont [1]{#1}%
\providecommand \href@noop [0]{\@secondoftwo}%
\providecommand \href [0]{\begingroup \@sanitize@url \@href}%
\providecommand \@href[1]{\@@startlink{#1}\@@href}%
\providecommand \@@href[1]{\endgroup#1\@@endlink}%
\providecommand \@sanitize@url [0]{\catcode `\\12\catcode `\$12\catcode
  `\&12\catcode `\#12\catcode `\^12\catcode `\_12\catcode `\%12\relax}%
\providecommand \@@startlink[1]{}%
\providecommand \@@endlink[0]{}%
\providecommand \url  [0]{\begingroup\@sanitize@url \@url }%
\providecommand \@url [1]{\endgroup\@href {#1}{\urlprefix }}%
\providecommand \urlprefix  [0]{URL }%
\providecommand \Eprint [0]{\href }%
\providecommand \doibase [0]{http://dx.doi.org/}%
\providecommand \selectlanguage [0]{\@gobble}%
\providecommand \bibinfo  [0]{\@secondoftwo}%
\providecommand \bibfield  [0]{\@secondoftwo}%
\providecommand \translation [1]{[#1]}%
\providecommand \BibitemOpen [0]{}%
\providecommand \bibitemStop [0]{}%
\providecommand \bibitemNoStop [0]{.\EOS\space}%
\providecommand \EOS [0]{\spacefactor3000\relax}%
\providecommand \BibitemShut  [1]{\csname bibitem#1\endcsname}%
\let\auto@bib@innerbib\@empty
\bibitem [{\citenamefont {Weinberg}(1967)}]{Weinberg:1967tq}%
  \BibitemOpen
  \bibfield  {author} {\bibinfo {author} {\bibfnamefont {S.}~\bibnamefont
  {Weinberg}},\ }\href {\doibase 10.1103/PhysRevLett.19.1264} {\bibfield
  {journal} {\bibinfo  {journal} {Phys. Rev. Lett.}\ }\textbf {\bibinfo
  {volume} {19}},\ \bibinfo {pages} {1264} (\bibinfo {year}
  {1967})}\BibitemShut {NoStop}%
\bibitem [{\citenamefont {Aad}\ \emph {et~al.}(2012)\citenamefont {Aad} \emph
  {et~al.}}]{ATLAS:2012yve}%
  \BibitemOpen
  \bibfield  {author} {\bibinfo {author} {\bibfnamefont {G.}~\bibnamefont
  {Aad}} \emph {et~al.} (\bibinfo {collaboration} {ATLAS}),\ }\href {\doibase
  10.1016/j.physletb.2012.08.020} {\bibfield  {journal} {\bibinfo  {journal}
  {Phys. Lett. B}\ }\textbf {\bibinfo {volume} {716}},\ \bibinfo {pages} {1}
  (\bibinfo {year} {2012})},\ \Eprint {http://arxiv.org/abs/1207.7214}
  {arXiv:1207.7214 [hep-ex]} \BibitemShut {NoStop}%
\bibitem [{\citenamefont {Aad}\ \emph {et~al.}(2015)\citenamefont {Aad} \emph
  {et~al.}}]{ATLAS:2015yey}%
  \BibitemOpen
  \bibfield  {author} {\bibinfo {author} {\bibfnamefont {G.}~\bibnamefont
  {Aad}} \emph {et~al.} (\bibinfo {collaboration} {ATLAS, CMS}),\ }\href
  {\doibase 10.1103/PhysRevLett.114.191803} {\bibfield  {journal} {\bibinfo
  {journal} {Phys. Rev. Lett.}\ }\textbf {\bibinfo {volume} {114}},\ \bibinfo
  {pages} {191803} (\bibinfo {year} {2015})},\ \Eprint
  {http://arxiv.org/abs/1503.07589} {arXiv:1503.07589 [hep-ex]} \BibitemShut
  {NoStop}%
\bibitem [{\citenamefont {Fukuda}\ \emph {et~al.}(1998)\citenamefont {Fukuda}
  \emph {et~al.}}]{Super-Kamiokande:1998kpq}%
  \BibitemOpen
  \bibfield  {author} {\bibinfo {author} {\bibfnamefont {Y.}~\bibnamefont
  {Fukuda}} \emph {et~al.} (\bibinfo {collaboration} {Super-Kamiokande}),\
  }\href {\doibase 10.1103/PhysRevLett.81.1562} {\bibfield  {journal} {\bibinfo
   {journal} {Phys. Rev. Lett.}\ }\textbf {\bibinfo {volume} {81}},\ \bibinfo
  {pages} {1562} (\bibinfo {year} {1998})},\ \Eprint
  {http://arxiv.org/abs/hep-ex/9807003} {arXiv:hep-ex/9807003} \BibitemShut
  {NoStop}%
\bibitem [{\citenamefont {Ahmad}\ \emph {et~al.}(2002)\citenamefont {Ahmad}
  \emph {et~al.}}]{SNO:2002tuh}%
  \BibitemOpen
  \bibfield  {author} {\bibinfo {author} {\bibfnamefont {Q.~R.}\ \bibnamefont
  {Ahmad}} \emph {et~al.} (\bibinfo {collaboration} {SNO}),\ }\href {\doibase
  10.1103/PhysRevLett.89.011301} {\bibfield  {journal} {\bibinfo  {journal}
  {Phys. Rev. Lett.}\ }\textbf {\bibinfo {volume} {89}},\ \bibinfo {pages}
  {011301} (\bibinfo {year} {2002})},\ \Eprint
  {http://arxiv.org/abs/nucl-ex/0204008} {arXiv:nucl-ex/0204008} \BibitemShut
  {NoStop}%
\bibitem [{\citenamefont {An}\ \emph {et~al.}(2012)\citenamefont {An} \emph
  {et~al.}}]{DayaBay:2012fng}%
  \BibitemOpen
  \bibfield  {author} {\bibinfo {author} {\bibfnamefont {F.~P.}\ \bibnamefont
  {An}} \emph {et~al.} (\bibinfo {collaboration} {Daya Bay}),\ }\href {\doibase
  10.1103/PhysRevLett.108.171803} {\bibfield  {journal} {\bibinfo  {journal}
  {Phys. Rev. Lett.}\ }\textbf {\bibinfo {volume} {108}},\ \bibinfo {pages}
  {171803} (\bibinfo {year} {2012})},\ \Eprint {http://arxiv.org/abs/1203.1669}
  {arXiv:1203.1669 [hep-ex]} \BibitemShut {NoStop}%
\bibitem [{\citenamefont {Bertone}\ \emph {et~al.}(2005)\citenamefont
  {Bertone}, \citenamefont {Hooper},\ and\ \citenamefont
  {Silk}}]{Bertone:2004pz}%
  \BibitemOpen
  \bibfield  {author} {\bibinfo {author} {\bibfnamefont {G.}~\bibnamefont
  {Bertone}}, \bibinfo {author} {\bibfnamefont {D.}~\bibnamefont {Hooper}}, \
  and\ \bibinfo {author} {\bibfnamefont {J.}~\bibnamefont {Silk}},\ }\href
  {\doibase 10.1016/j.physrep.2004.08.031} {\bibfield  {journal} {\bibinfo
  {journal} {Phys. Rept.}\ }\textbf {\bibinfo {volume} {405}},\ \bibinfo
  {pages} {279} (\bibinfo {year} {2005})},\ \Eprint
  {http://arxiv.org/abs/hep-ph/0404175} {arXiv:hep-ph/0404175} \BibitemShut
  {NoStop}%
\bibitem [{\citenamefont {Dine}\ and\ \citenamefont
  {Kusenko}(2003)}]{Dine:2003ax}%
  \BibitemOpen
  \bibfield  {author} {\bibinfo {author} {\bibfnamefont {M.}~\bibnamefont
  {Dine}}\ and\ \bibinfo {author} {\bibfnamefont {A.}~\bibnamefont {Kusenko}},\
  }\href {\doibase 10.1103/RevModPhys.76.1} {\bibfield  {journal} {\bibinfo
  {journal} {Rev. Mod. Phys.}\ }\textbf {\bibinfo {volume} {76}},\ \bibinfo
  {pages} {1} (\bibinfo {year} {2003})},\ \Eprint
  {http://arxiv.org/abs/hep-ph/0303065} {arXiv:hep-ph/0303065} \BibitemShut
  {NoStop}%
\bibitem [{\citenamefont {Aghanim}\ \emph {et~al.}(2020)\citenamefont {Aghanim}
  \emph {et~al.}}]{Planck:2018vyg}%
  \BibitemOpen
  \bibfield  {author} {\bibinfo {author} {\bibfnamefont {N.}~\bibnamefont
  {Aghanim}} \emph {et~al.} (\bibinfo {collaboration} {Planck}),\ }\href
  {\doibase 10.1051/0004-6361/201833910} {\bibfield  {journal} {\bibinfo
  {journal} {Astron. Astrophys.}\ }\textbf {\bibinfo {volume} {641}},\ \bibinfo
  {pages} {A6} (\bibinfo {year} {2020})},\ \bibinfo {note} {[Erratum:
  Astron.Astrophys. 652, C4 (2021)]},\ \Eprint
  {http://arxiv.org/abs/1807.06209} {arXiv:1807.06209 [astro-ph.CO]}
  \BibitemShut {NoStop}%
\bibitem [{\citenamefont {Sakharov}(1967)}]{Sakharov:1967dj}%
  \BibitemOpen
  \bibfield  {author} {\bibinfo {author} {\bibfnamefont {A.~D.}\ \bibnamefont
  {Sakharov}},\ }\href {\doibase 10.1070/PU1991v034n05ABEH002497} {\bibfield
  {journal} {\bibinfo  {journal} {Pisma Zh. Eksp. Teor. Fiz.}\ }\textbf
  {\bibinfo {volume} {5}},\ \bibinfo {pages} {32} (\bibinfo {year}
  {1967})}\BibitemShut {NoStop}%
\bibitem [{\citenamefont {Fukugita}\ and\ \citenamefont
  {Yanagida}(1986)}]{Fukugita:1986hr}%
  \BibitemOpen
  \bibfield  {author} {\bibinfo {author} {\bibfnamefont {M.}~\bibnamefont
  {Fukugita}}\ and\ \bibinfo {author} {\bibfnamefont {T.}~\bibnamefont
  {Yanagida}},\ }\href {\doibase 10.1016/0370-2693(86)91126-3} {\bibfield
  {journal} {\bibinfo  {journal} {Phys. Lett. B}\ }\textbf {\bibinfo {volume}
  {174}},\ \bibinfo {pages} {45} (\bibinfo {year} {1986})}\BibitemShut
  {NoStop}%
\bibitem [{\citenamefont {Cohen}\ \emph {et~al.}(1993)\citenamefont {Cohen},
  \citenamefont {Kaplan},\ and\ \citenamefont {Nelson}}]{Cohen:1993nk}%
  \BibitemOpen
  \bibfield  {author} {\bibinfo {author} {\bibfnamefont {A.~G.}\ \bibnamefont
  {Cohen}}, \bibinfo {author} {\bibfnamefont {D.~B.}\ \bibnamefont {Kaplan}}, \
  and\ \bibinfo {author} {\bibfnamefont {A.~E.}\ \bibnamefont {Nelson}},\
  }\href {\doibase 10.1146/annurev.ns.43.120193.000331} {\bibfield  {journal}
  {\bibinfo  {journal} {Ann. Rev. Nucl. Part. Sci.}\ }\textbf {\bibinfo
  {volume} {43}},\ \bibinfo {pages} {27} (\bibinfo {year} {1993})},\ \Eprint
  {http://arxiv.org/abs/hep-ph/9302210} {arXiv:hep-ph/9302210} \BibitemShut
  {NoStop}%
\bibitem [{\citenamefont {Trodden}(1999)}]{Trodden:1998ym}%
  \BibitemOpen
  \bibfield  {author} {\bibinfo {author} {\bibfnamefont {M.}~\bibnamefont
  {Trodden}},\ }\href {\doibase 10.1103/RevModPhys.71.1463} {\bibfield
  {journal} {\bibinfo  {journal} {Rev. Mod. Phys.}\ }\textbf {\bibinfo {volume}
  {71}},\ \bibinfo {pages} {1463} (\bibinfo {year} {1999})},\ \Eprint
  {http://arxiv.org/abs/hep-ph/9803479} {arXiv:hep-ph/9803479} \BibitemShut
  {NoStop}%
\bibitem [{\citenamefont {Morrissey}\ and\ \citenamefont
  {Ramsey-Musolf}(2012)}]{Morrissey:2012db}%
  \BibitemOpen
  \bibfield  {author} {\bibinfo {author} {\bibfnamefont {D.~E.}\ \bibnamefont
  {Morrissey}}\ and\ \bibinfo {author} {\bibfnamefont {M.~J.}\ \bibnamefont
  {Ramsey-Musolf}},\ }\href {\doibase 10.1088/1367-2630/14/12/125003}
  {\bibfield  {journal} {\bibinfo  {journal} {New J. Phys.}\ }\textbf {\bibinfo
  {volume} {14}},\ \bibinfo {pages} {125003} (\bibinfo {year} {2012})},\
  \Eprint {http://arxiv.org/abs/1206.2942} {arXiv:1206.2942 [hep-ph]}
  \BibitemShut {NoStop}%
\bibitem [{\citenamefont {Affleck}\ and\ \citenamefont
  {Dine}(1985)}]{Affleck:1984fy}%
  \BibitemOpen
  \bibfield  {author} {\bibinfo {author} {\bibfnamefont {I.}~\bibnamefont
  {Affleck}}\ and\ \bibinfo {author} {\bibfnamefont {M.}~\bibnamefont {Dine}},\
  }\href {\doibase 10.1016/0550-3213(85)90021-5} {\bibfield  {journal}
  {\bibinfo  {journal} {Nucl. Phys. B}\ }\textbf {\bibinfo {volume} {249}},\
  \bibinfo {pages} {361} (\bibinfo {year} {1985})}\BibitemShut {NoStop}%
\bibitem [{\citenamefont {Cohen}\ and\ \citenamefont
  {Kaplan}(1988)}]{Cohen:1988kt}%
  \BibitemOpen
  \bibfield  {author} {\bibinfo {author} {\bibfnamefont {A.~G.}\ \bibnamefont
  {Cohen}}\ and\ \bibinfo {author} {\bibfnamefont {D.~B.}\ \bibnamefont
  {Kaplan}},\ }\href {\doibase 10.1016/0550-3213(88)90134-4} {\bibfield
  {journal} {\bibinfo  {journal} {Nucl. Phys. B}\ }\textbf {\bibinfo {volume}
  {308}},\ \bibinfo {pages} {913} (\bibinfo {year} {1988})}\BibitemShut
  {NoStop}%
\bibitem [{\citenamefont {Li}\ \emph {et~al.}(2010)\citenamefont {Li},
  \citenamefont {Profumo},\ and\ \citenamefont {Ramsey-Musolf}}]{Li:2010ax}%
  \BibitemOpen
  \bibfield  {author} {\bibinfo {author} {\bibfnamefont {Y.}~\bibnamefont
  {Li}}, \bibinfo {author} {\bibfnamefont {S.}~\bibnamefont {Profumo}}, \ and\
  \bibinfo {author} {\bibfnamefont {M.}~\bibnamefont {Ramsey-Musolf}},\ }\href
  {\doibase 10.1007/JHEP08(2010)062} {\bibfield  {journal} {\bibinfo  {journal}
  {JHEP}\ }\textbf {\bibinfo {volume} {08}},\ \bibinfo {pages} {062} (\bibinfo
  {year} {2010})},\ \Eprint {http://arxiv.org/abs/1006.1440} {arXiv:1006.1440
  [hep-ph]} \BibitemShut {NoStop}%
\bibitem [{\citenamefont {Chao}(2019)}]{Chao:2017oux}%
  \BibitemOpen
  \bibfield  {author} {\bibinfo {author} {\bibfnamefont {W.}~\bibnamefont
  {Chao}},\ }\href {\doibase 10.1016/j.physletb.2019.07.025} {\bibfield
  {journal} {\bibinfo  {journal} {Phys. Lett. B}\ }\textbf {\bibinfo {volume}
  {796}},\ \bibinfo {pages} {102} (\bibinfo {year} {2019})},\ \Eprint
  {http://arxiv.org/abs/1706.01041} {arXiv:1706.01041 [hep-ph]} \BibitemShut
  {NoStop}%
\bibitem [{\citenamefont {Co}\ and\ \citenamefont
  {Harigaya}(2020)}]{Co:2019wyp}%
  \BibitemOpen
  \bibfield  {author} {\bibinfo {author} {\bibfnamefont {R.~T.}\ \bibnamefont
  {Co}}\ and\ \bibinfo {author} {\bibfnamefont {K.}~\bibnamefont {Harigaya}},\
  }\href {\doibase 10.1103/PhysRevLett.124.111602} {\bibfield  {journal}
  {\bibinfo  {journal} {Phys. Rev. Lett.}\ }\textbf {\bibinfo {volume} {124}},\
  \bibinfo {pages} {111602} (\bibinfo {year} {2020})},\ \Eprint
  {http://arxiv.org/abs/1910.02080} {arXiv:1910.02080 [hep-ph]} \BibitemShut
  {NoStop}%
\bibitem [{\citenamefont {Domcke}\ \emph {et~al.}(2020)\citenamefont {Domcke},
  \citenamefont {Ema}, \citenamefont {Mukaida},\ and\ \citenamefont
  {Yamada}}]{Domcke:2020kcp}%
  \BibitemOpen
  \bibfield  {author} {\bibinfo {author} {\bibfnamefont {V.}~\bibnamefont
  {Domcke}}, \bibinfo {author} {\bibfnamefont {Y.}~\bibnamefont {Ema}},
  \bibinfo {author} {\bibfnamefont {K.}~\bibnamefont {Mukaida}}, \ and\
  \bibinfo {author} {\bibfnamefont {M.}~\bibnamefont {Yamada}},\ }\href
  {\doibase 10.1007/JHEP08(2020)096} {\bibfield  {journal} {\bibinfo  {journal}
  {JHEP}\ }\textbf {\bibinfo {volume} {08}},\ \bibinfo {pages} {096} (\bibinfo
  {year} {2020})},\ \Eprint {http://arxiv.org/abs/2006.03148} {arXiv:2006.03148
  [hep-ph]} \BibitemShut {NoStop}%
\bibitem [{\citenamefont {Domcke}\ \emph {et~al.}(2019)\citenamefont {Domcke},
  \citenamefont {von Harling}, \citenamefont {Morgante},\ and\ \citenamefont
  {Mukaida}}]{Domcke:2019mnd}%
  \BibitemOpen
  \bibfield  {author} {\bibinfo {author} {\bibfnamefont {V.}~\bibnamefont
  {Domcke}}, \bibinfo {author} {\bibfnamefont {B.}~\bibnamefont {von Harling}},
  \bibinfo {author} {\bibfnamefont {E.}~\bibnamefont {Morgante}}, \ and\
  \bibinfo {author} {\bibfnamefont {K.}~\bibnamefont {Mukaida}},\ }\href
  {\doibase 10.1088/1475-7516/2019/10/032} {\bibfield  {journal} {\bibinfo
  {journal} {JCAP}\ }\textbf {\bibinfo {volume} {10}},\ \bibinfo {pages} {032}
  (\bibinfo {year} {2019})},\ \Eprint {http://arxiv.org/abs/1905.13318}
  {arXiv:1905.13318 [hep-ph]} \BibitemShut {NoStop}%
\bibitem [{\citenamefont {Croon}\ \emph {et~al.}(2020)\citenamefont {Croon},
  \citenamefont {Howard}, \citenamefont {Ipek},\ and\ \citenamefont
  {Tait}}]{Croon:2019ugf}%
  \BibitemOpen
  \bibfield  {author} {\bibinfo {author} {\bibfnamefont {D.}~\bibnamefont
  {Croon}}, \bibinfo {author} {\bibfnamefont {J.~N.}\ \bibnamefont {Howard}},
  \bibinfo {author} {\bibfnamefont {S.}~\bibnamefont {Ipek}}, \ and\ \bibinfo
  {author} {\bibfnamefont {T.~M.~P.}\ \bibnamefont {Tait}},\ }\href {\doibase
  10.1103/PhysRevD.101.055042} {\bibfield  {journal} {\bibinfo  {journal}
  {Phys. Rev. D}\ }\textbf {\bibinfo {volume} {101}},\ \bibinfo {pages}
  {055042} (\bibinfo {year} {2020})},\ \Eprint
  {http://arxiv.org/abs/1911.01432} {arXiv:1911.01432 [hep-ph]} \BibitemShut
  {NoStop}%
\bibitem [{\citenamefont {Domcke}\ \emph {et~al.}(2021)\citenamefont {Domcke},
  \citenamefont {Kamada}, \citenamefont {Mukaida}, \citenamefont {Schmitz},\
  and\ \citenamefont {Yamada}}]{Domcke:2020quw}%
  \BibitemOpen
  \bibfield  {author} {\bibinfo {author} {\bibfnamefont {V.}~\bibnamefont
  {Domcke}}, \bibinfo {author} {\bibfnamefont {K.}~\bibnamefont {Kamada}},
  \bibinfo {author} {\bibfnamefont {K.}~\bibnamefont {Mukaida}}, \bibinfo
  {author} {\bibfnamefont {K.}~\bibnamefont {Schmitz}}, \ and\ \bibinfo
  {author} {\bibfnamefont {M.}~\bibnamefont {Yamada}},\ }\href {\doibase
  10.1103/PhysRevLett.126.201802} {\bibfield  {journal} {\bibinfo  {journal}
  {Phys. Rev. Lett.}\ }\textbf {\bibinfo {volume} {126}},\ \bibinfo {pages}
  {201802} (\bibinfo {year} {2021})},\ \Eprint
  {http://arxiv.org/abs/2011.09347} {arXiv:2011.09347 [hep-ph]} \BibitemShut
  {NoStop}%
\bibitem [{\citenamefont {Marshak}\ and\ \citenamefont
  {Mohapatra}(1980)}]{Marshak:1979fm}%
  \BibitemOpen
  \bibfield  {author} {\bibinfo {author} {\bibfnamefont {R.~E.}\ \bibnamefont
  {Marshak}}\ and\ \bibinfo {author} {\bibfnamefont {R.~N.}\ \bibnamefont
  {Mohapatra}},\ }\href {\doibase 10.1016/0370-2693(80)90436-0} {\bibfield
  {journal} {\bibinfo  {journal} {Phys. Lett. B}\ }\textbf {\bibinfo {volume}
  {91}},\ \bibinfo {pages} {222} (\bibinfo {year} {1980})}\BibitemShut
  {NoStop}%
\bibitem [{\citenamefont {Mukaida}\ \emph {et~al.}(2022)\citenamefont
  {Mukaida}, \citenamefont {Schmitz},\ and\ \citenamefont
  {Yamada}}]{Mukaida:2021sgv}%
  \BibitemOpen
  \bibfield  {author} {\bibinfo {author} {\bibfnamefont {K.}~\bibnamefont
  {Mukaida}}, \bibinfo {author} {\bibfnamefont {K.}~\bibnamefont {Schmitz}}, \
  and\ \bibinfo {author} {\bibfnamefont {M.}~\bibnamefont {Yamada}},\ }\href
  {\doibase 10.1103/PhysRevLett.129.011803} {\bibfield  {journal} {\bibinfo
  {journal} {Phys. Rev. Lett.}\ }\textbf {\bibinfo {volume} {129}},\ \bibinfo
  {pages} {011803} (\bibinfo {year} {2022})},\ \Eprint
  {http://arxiv.org/abs/2111.03082} {arXiv:2111.03082 [hep-ph]} \BibitemShut
  {NoStop}%
\bibitem [{\citenamefont {Davoudiasl}\ \emph {et~al.}(2010)\citenamefont
  {Davoudiasl}, \citenamefont {Morrissey}, \citenamefont {Sigurdson},\ and\
  \citenamefont {Tulin}}]{Davoudiasl:2010am}%
  \BibitemOpen
  \bibfield  {author} {\bibinfo {author} {\bibfnamefont {H.}~\bibnamefont
  {Davoudiasl}}, \bibinfo {author} {\bibfnamefont {D.~E.}\ \bibnamefont
  {Morrissey}}, \bibinfo {author} {\bibfnamefont {K.}~\bibnamefont
  {Sigurdson}}, \ and\ \bibinfo {author} {\bibfnamefont {S.}~\bibnamefont
  {Tulin}},\ }\href {\doibase 10.1103/PhysRevLett.105.211304} {\bibfield
  {journal} {\bibinfo  {journal} {Phys. Rev. Lett.}\ }\textbf {\bibinfo
  {volume} {105}},\ \bibinfo {pages} {211304} (\bibinfo {year} {2010})},\
  \Eprint {http://arxiv.org/abs/1008.2399} {arXiv:1008.2399 [hep-ph]}
  \BibitemShut {NoStop}%
\bibitem [{\citenamefont {Shelton}\ and\ \citenamefont
  {Zurek}(2010)}]{Shelton:2010ta}%
  \BibitemOpen
  \bibfield  {author} {\bibinfo {author} {\bibfnamefont {J.}~\bibnamefont
  {Shelton}}\ and\ \bibinfo {author} {\bibfnamefont {K.~M.}\ \bibnamefont
  {Zurek}},\ }\href {\doibase 10.1103/PhysRevD.82.123512} {\bibfield  {journal}
  {\bibinfo  {journal} {Phys. Rev. D}\ }\textbf {\bibinfo {volume} {82}},\
  \bibinfo {pages} {123512} (\bibinfo {year} {2010})},\ \Eprint
  {http://arxiv.org/abs/1008.1997} {arXiv:1008.1997 [hep-ph]} \BibitemShut
  {NoStop}%
\bibitem [{\citenamefont {Cui}\ \emph {et~al.}(2012)\citenamefont {Cui},
  \citenamefont {Randall},\ and\ \citenamefont {Shuve}}]{Cui:2011ab}%
  \BibitemOpen
  \bibfield  {author} {\bibinfo {author} {\bibfnamefont {Y.}~\bibnamefont
  {Cui}}, \bibinfo {author} {\bibfnamefont {L.}~\bibnamefont {Randall}}, \ and\
  \bibinfo {author} {\bibfnamefont {B.}~\bibnamefont {Shuve}},\ }\href
  {\doibase 10.1007/JHEP04(2012)075} {\bibfield  {journal} {\bibinfo  {journal}
  {JHEP}\ }\textbf {\bibinfo {volume} {04}},\ \bibinfo {pages} {075} (\bibinfo
  {year} {2012})},\ \Eprint {http://arxiv.org/abs/1112.2704} {arXiv:1112.2704
  [hep-ph]} \BibitemShut {NoStop}%
\bibitem [{\citenamefont {Hall}\ \emph {et~al.}(2010)\citenamefont {Hall},
  \citenamefont {March-Russell},\ and\ \citenamefont {West}}]{Hall:2010jx}%
  \BibitemOpen
  \bibfield  {author} {\bibinfo {author} {\bibfnamefont {L.~J.}\ \bibnamefont
  {Hall}}, \bibinfo {author} {\bibfnamefont {J.}~\bibnamefont {March-Russell}},
  \ and\ \bibinfo {author} {\bibfnamefont {S.~M.}\ \bibnamefont {West}},\
  }\href@noop {} {\  (\bibinfo {year} {2010})},\ \Eprint
  {http://arxiv.org/abs/1010.0245} {arXiv:1010.0245 [hep-ph]} \BibitemShut
  {NoStop}%
\bibitem [{\citenamefont {Elor}\ \emph {et~al.}(2019)\citenamefont {Elor},
  \citenamefont {Escudero},\ and\ \citenamefont {Nelson}}]{Elor:2018twp}%
  \BibitemOpen
  \bibfield  {author} {\bibinfo {author} {\bibfnamefont {G.}~\bibnamefont
  {Elor}}, \bibinfo {author} {\bibfnamefont {M.}~\bibnamefont {Escudero}}, \
  and\ \bibinfo {author} {\bibfnamefont {A.}~\bibnamefont {Nelson}},\ }\href
  {\doibase 10.1103/PhysRevD.99.035031} {\bibfield  {journal} {\bibinfo
  {journal} {Phys. Rev. D}\ }\textbf {\bibinfo {volume} {99}},\ \bibinfo
  {pages} {035031} (\bibinfo {year} {2019})},\ \Eprint
  {http://arxiv.org/abs/1810.00880} {arXiv:1810.00880 [hep-ph]} \BibitemShut
  {NoStop}%
\bibitem [{\citenamefont {Chao}\ and\ \citenamefont
  {Peng}(2023)}]{Chao:2023ojl}%
  \BibitemOpen
  \bibfield  {author} {\bibinfo {author} {\bibfnamefont {W.}~\bibnamefont
  {Chao}}\ and\ \bibinfo {author} {\bibfnamefont {Y.-Q.}\ \bibnamefont
  {Peng}},\ }\href@noop {} {\  (\bibinfo {year} {2023})},\ \Eprint
  {http://arxiv.org/abs/2311.06469} {arXiv:2311.06469 [hep-ph]} \BibitemShut
  {NoStop}%
\bibitem [{\citenamefont {Elor}\ \emph {et~al.}(2022)\citenamefont {Elor} \emph
  {et~al.}}]{Elor:2022hpa}%
  \BibitemOpen
  \bibfield  {author} {\bibinfo {author} {\bibfnamefont {G.}~\bibnamefont
  {Elor}} \emph {et~al.},\ }in\ \href@noop {} {\emph {\bibinfo {booktitle}
  {{Snowmass 2021}}}}\ (\bibinfo {year} {2022})\ \Eprint
  {http://arxiv.org/abs/2203.05010} {arXiv:2203.05010 [hep-ph]} \BibitemShut
  {NoStop}%
\bibitem [{\citenamefont {Freese}\ \emph {et~al.}(1990)\citenamefont {Freese},
  \citenamefont {Frieman},\ and\ \citenamefont {Olinto}}]{Freese:1990rb}%
  \BibitemOpen
  \bibfield  {author} {\bibinfo {author} {\bibfnamefont {K.}~\bibnamefont
  {Freese}}, \bibinfo {author} {\bibfnamefont {J.~A.}\ \bibnamefont {Frieman}},
  \ and\ \bibinfo {author} {\bibfnamefont {A.~V.}\ \bibnamefont {Olinto}},\
  }\href {\doibase 10.1103/PhysRevLett.65.3233} {\bibfield  {journal} {\bibinfo
   {journal} {Phys. Rev. Lett.}\ }\textbf {\bibinfo {volume} {65}},\ \bibinfo
  {pages} {3233} (\bibinfo {year} {1990})}\BibitemShut {NoStop}%
\bibitem [{\citenamefont {Alexander}\ \emph {et~al.}(2006)\citenamefont
  {Alexander}, \citenamefont {Peskin},\ and\ \citenamefont
  {Sheikh-Jabbari}}]{Alexander:2004us}%
  \BibitemOpen
  \bibfield  {author} {\bibinfo {author} {\bibfnamefont {S.~H.-S.}\
  \bibnamefont {Alexander}}, \bibinfo {author} {\bibfnamefont {M.~E.}\
  \bibnamefont {Peskin}}, \ and\ \bibinfo {author} {\bibfnamefont {M.~M.}\
  \bibnamefont {Sheikh-Jabbari}},\ }\href {\doibase
  10.1103/PhysRevLett.96.081301} {\bibfield  {journal} {\bibinfo  {journal}
  {Phys. Rev. Lett.}\ }\textbf {\bibinfo {volume} {96}},\ \bibinfo {pages}
  {081301} (\bibinfo {year} {2006})},\ \Eprint
  {http://arxiv.org/abs/hep-th/0403069} {arXiv:hep-th/0403069} \BibitemShut
  {NoStop}%
\bibitem [{\citenamefont {Akita}\ \emph {et~al.}(2017)\citenamefont {Akita},
  \citenamefont {Kobayashi},\ and\ \citenamefont {Otsuka}}]{Akita:2017ecc}%
  \BibitemOpen
  \bibfield  {author} {\bibinfo {author} {\bibfnamefont {K.}~\bibnamefont
  {Akita}}, \bibinfo {author} {\bibfnamefont {T.}~\bibnamefont {Kobayashi}}, \
  and\ \bibinfo {author} {\bibfnamefont {H.}~\bibnamefont {Otsuka}},\ }\href
  {\doibase 10.1088/1475-7516/2017/04/042} {\bibfield  {journal} {\bibinfo
  {journal} {JCAP}\ }\textbf {\bibinfo {volume} {04}},\ \bibinfo {pages} {042}
  (\bibinfo {year} {2017})},\ \Eprint {http://arxiv.org/abs/1702.01604}
  {arXiv:1702.01604 [hep-ph]} \BibitemShut {NoStop}%
\bibitem [{\citenamefont {Maleknejad}(2021)}]{Maleknejad:2020yys}%
  \BibitemOpen
  \bibfield  {author} {\bibinfo {author} {\bibfnamefont {A.}~\bibnamefont
  {Maleknejad}},\ }\href {\doibase 10.1103/PhysRevD.104.083518} {\bibfield
  {journal} {\bibinfo  {journal} {Phys. Rev. D}\ }\textbf {\bibinfo {volume}
  {104}},\ \bibinfo {pages} {083518} (\bibinfo {year} {2021})},\ \Eprint
  {http://arxiv.org/abs/2012.11516} {arXiv:2012.11516 [hep-ph]} \BibitemShut
  {NoStop}%
\bibitem [{\citenamefont {Maleknejad}(2020)}]{Maleknejad:2020pec}%
  \BibitemOpen
  \bibfield  {author} {\bibinfo {author} {\bibfnamefont {A.}~\bibnamefont
  {Maleknejad}},\ }\href {\doibase 10.1007/JHEP06(2021)113} {\bibfield
  {journal} {\bibinfo  {journal} {JHEP}\ }\textbf {\bibinfo {volume} {21}},\
  \bibinfo {pages} {113} (\bibinfo {year} {2020})},\ \Eprint
  {http://arxiv.org/abs/2103.14611} {arXiv:2103.14611 [hep-ph]} \BibitemShut
  {NoStop}%
\bibitem [{\citenamefont {Domcke}\ \emph {et~al.}(2023)\citenamefont {Domcke},
  \citenamefont {Kamada}, \citenamefont {Mukaida}, \citenamefont {Schmitz},\
  and\ \citenamefont {Yamada}}]{Domcke:2022kfs}%
  \BibitemOpen
  \bibfield  {author} {\bibinfo {author} {\bibfnamefont {V.}~\bibnamefont
  {Domcke}}, \bibinfo {author} {\bibfnamefont {K.}~\bibnamefont {Kamada}},
  \bibinfo {author} {\bibfnamefont {K.}~\bibnamefont {Mukaida}}, \bibinfo
  {author} {\bibfnamefont {K.}~\bibnamefont {Schmitz}}, \ and\ \bibinfo
  {author} {\bibfnamefont {M.}~\bibnamefont {Yamada}},\ }\href {\doibase
  10.1007/JHEP01(2023)053} {\bibfield  {journal} {\bibinfo  {journal} {JHEP}\
  }\textbf {\bibinfo {volume} {01}},\ \bibinfo {pages} {053} (\bibinfo {year}
  {2023})},\ \Eprint {http://arxiv.org/abs/2210.06412} {arXiv:2210.06412
  [hep-ph]} \BibitemShut {NoStop}%
\bibitem [{\citenamefont {Hashiba}\ \emph {et~al.}(2022)\citenamefont
  {Hashiba}, \citenamefont {Kamada},\ and\ \citenamefont
  {Nakatsuka}}]{Hashiba:2021gmn}%
  \BibitemOpen
  \bibfield  {author} {\bibinfo {author} {\bibfnamefont {S.}~\bibnamefont
  {Hashiba}}, \bibinfo {author} {\bibfnamefont {K.}~\bibnamefont {Kamada}}, \
  and\ \bibinfo {author} {\bibfnamefont {H.}~\bibnamefont {Nakatsuka}},\ }\href
  {\doibase 10.1088/1475-7516/2022/04/058} {\bibfield  {journal} {\bibinfo
  {journal} {JCAP}\ }\textbf {\bibinfo {volume} {04}},\ \bibinfo {pages} {058}
  (\bibinfo {year} {2022})},\ \Eprint {http://arxiv.org/abs/2110.10822}
  {arXiv:2110.10822 [hep-ph]} \BibitemShut {NoStop}%
\bibitem [{\citenamefont {Tishue}\ and\ \citenamefont
  {Caldwell}(2021)}]{Tishue:2021blv}%
  \BibitemOpen
  \bibfield  {author} {\bibinfo {author} {\bibfnamefont {A.~J.}\ \bibnamefont
  {Tishue}}\ and\ \bibinfo {author} {\bibfnamefont {R.~R.}\ \bibnamefont
  {Caldwell}},\ }\href {\doibase 10.1103/PhysRevD.104.063531} {\bibfield
  {journal} {\bibinfo  {journal} {Phys. Rev. D}\ }\textbf {\bibinfo {volume}
  {104}},\ \bibinfo {pages} {063531} (\bibinfo {year} {2021})},\ \Eprint
  {http://arxiv.org/abs/2105.08073} {arXiv:2105.08073 [astro-ph.CO]}
  \BibitemShut {NoStop}%
\bibitem [{\citenamefont {Cado}\ \emph {et~al.}(2021)\citenamefont {Cado},
  \citenamefont {von Harling}, \citenamefont {Mass\'o},\ and\ \citenamefont
  {Quir\'os}}]{Cado:2021bia}%
  \BibitemOpen
  \bibfield  {author} {\bibinfo {author} {\bibfnamefont {Y.}~\bibnamefont
  {Cado}}, \bibinfo {author} {\bibfnamefont {B.}~\bibnamefont {von Harling}},
  \bibinfo {author} {\bibfnamefont {E.}~\bibnamefont {Mass\'o}}, \ and\
  \bibinfo {author} {\bibfnamefont {M.}~\bibnamefont {Quir\'os}},\ }\href
  {\doibase 10.1088/1475-7516/2021/07/049} {\bibfield  {journal} {\bibinfo
  {journal} {JCAP}\ }\textbf {\bibinfo {volume} {07}},\ \bibinfo {pages} {049}
  (\bibinfo {year} {2021})},\ \Eprint {http://arxiv.org/abs/2102.13650}
  {arXiv:2102.13650 [hep-ph]} \BibitemShut {NoStop}%
\bibitem [{\citenamefont {Kamada}\ \emph {et~al.}(2020)\citenamefont {Kamada},
  \citenamefont {Kume}, \citenamefont {Yamada},\ and\ \citenamefont
  {Yokoyama}}]{Kamada:2019ewe}%
  \BibitemOpen
  \bibfield  {author} {\bibinfo {author} {\bibfnamefont {K.}~\bibnamefont
  {Kamada}}, \bibinfo {author} {\bibfnamefont {J.}~\bibnamefont {Kume}},
  \bibinfo {author} {\bibfnamefont {Y.}~\bibnamefont {Yamada}}, \ and\ \bibinfo
  {author} {\bibfnamefont {J.}~\bibnamefont {Yokoyama}},\ }\href {\doibase
  10.1088/1475-7516/2020/01/016} {\bibfield  {journal} {\bibinfo  {journal}
  {JCAP}\ }\textbf {\bibinfo {volume} {01}},\ \bibinfo {pages} {016} (\bibinfo
  {year} {2020})},\ \Eprint {http://arxiv.org/abs/1911.02657} {arXiv:1911.02657
  [hep-ph]} \BibitemShut {NoStop}%
\bibitem [{\citenamefont {Jim\'enez}\ \emph {et~al.}(2017)\citenamefont
  {Jim\'enez}, \citenamefont {Kamada}, \citenamefont {Schmitz},\ and\
  \citenamefont {Xu}}]{Jimenez:2017cdr}%
  \BibitemOpen
  \bibfield  {author} {\bibinfo {author} {\bibfnamefont {D.}~\bibnamefont
  {Jim\'enez}}, \bibinfo {author} {\bibfnamefont {K.}~\bibnamefont {Kamada}},
  \bibinfo {author} {\bibfnamefont {K.}~\bibnamefont {Schmitz}}, \ and\
  \bibinfo {author} {\bibfnamefont {X.-J.}\ \bibnamefont {Xu}},\ }\href
  {\doibase 10.1088/1475-7516/2017/12/011} {\bibfield  {journal} {\bibinfo
  {journal} {JCAP}\ }\textbf {\bibinfo {volume} {12}},\ \bibinfo {pages} {011}
  (\bibinfo {year} {2017})},\ \Eprint {http://arxiv.org/abs/1707.07943}
  {arXiv:1707.07943 [hep-ph]} \BibitemShut {NoStop}%
\bibitem [{\citenamefont {Fileviez~Perez}\ and\ \citenamefont
  {Wise}(2010)}]{FileviezPerez:2010gw}%
  \BibitemOpen
  \bibfield  {author} {\bibinfo {author} {\bibfnamefont {P.}~\bibnamefont
  {Fileviez~Perez}}\ and\ \bibinfo {author} {\bibfnamefont {M.~B.}\
  \bibnamefont {Wise}},\ }\href {\doibase 10.1103/PhysRevD.82.079901}
  {\bibfield  {journal} {\bibinfo  {journal} {Phys. Rev. D}\ }\textbf {\bibinfo
  {volume} {82}},\ \bibinfo {pages} {011901} (\bibinfo {year} {2010})},\
  \bibinfo {note} {[Erratum: Phys.Rev.D 82, 079901 (2010)]},\ \Eprint
  {http://arxiv.org/abs/1002.1754} {arXiv:1002.1754 [hep-ph]} \BibitemShut
  {NoStop}%
\bibitem [{\citenamefont {Dulaney}\ \emph {et~al.}(2011)\citenamefont
  {Dulaney}, \citenamefont {Fileviez~Perez},\ and\ \citenamefont
  {Wise}}]{Dulaney:2010dj}%
  \BibitemOpen
  \bibfield  {author} {\bibinfo {author} {\bibfnamefont {T.~R.}\ \bibnamefont
  {Dulaney}}, \bibinfo {author} {\bibfnamefont {P.}~\bibnamefont
  {Fileviez~Perez}}, \ and\ \bibinfo {author} {\bibfnamefont {M.~B.}\
  \bibnamefont {Wise}},\ }\href {\doibase 10.1103/PhysRevD.83.023520}
  {\bibfield  {journal} {\bibinfo  {journal} {Phys. Rev. D}\ }\textbf {\bibinfo
  {volume} {83}},\ \bibinfo {pages} {023520} (\bibinfo {year} {2011})},\
  \Eprint {http://arxiv.org/abs/1005.0617} {arXiv:1005.0617 [hep-ph]}
  \BibitemShut {NoStop}%
\bibitem [{\citenamefont {Chao}(2011)}]{Chao:2010mp}%
  \BibitemOpen
  \bibfield  {author} {\bibinfo {author} {\bibfnamefont {W.}~\bibnamefont
  {Chao}},\ }\href {\doibase 10.1016/j.physletb.2010.10.056} {\bibfield
  {journal} {\bibinfo  {journal} {Phys. Lett. B}\ }\textbf {\bibinfo {volume}
  {695}},\ \bibinfo {pages} {157} (\bibinfo {year} {2011})},\ \Eprint
  {http://arxiv.org/abs/1005.1024} {arXiv:1005.1024 [hep-ph]} \BibitemShut
  {NoStop}%
\bibitem [{\citenamefont {Langacker}(2009)}]{Langacker:2008yv}%
  \BibitemOpen
  \bibfield  {author} {\bibinfo {author} {\bibfnamefont {P.}~\bibnamefont
  {Langacker}},\ }\href {\doibase 10.1103/RevModPhys.81.1199} {\bibfield
  {journal} {\bibinfo  {journal} {Rev. Mod. Phys.}\ }\textbf {\bibinfo {volume}
  {81}},\ \bibinfo {pages} {1199} (\bibinfo {year} {2009})},\ \Eprint
  {http://arxiv.org/abs/0801.1345} {arXiv:0801.1345 [hep-ph]} \BibitemShut
  {NoStop}%
\bibitem [{\citenamefont {Chao}(2018)}]{Chao:2017rwv}%
  \BibitemOpen
  \bibfield  {author} {\bibinfo {author} {\bibfnamefont {W.}~\bibnamefont
  {Chao}},\ }\href {\doibase 10.1140/epjc/s10052-018-5547-0} {\bibfield
  {journal} {\bibinfo  {journal} {Eur. Phys. J. C}\ }\textbf {\bibinfo {volume}
  {78}},\ \bibinfo {pages} {103} (\bibinfo {year} {2018})},\ \Eprint
  {http://arxiv.org/abs/1707.07858} {arXiv:1707.07858 [hep-ph]} \BibitemShut
  {NoStop}%
\bibitem [{\citenamefont {Mohapatra}\ and\ \citenamefont
  {Marshak}(1980)}]{Mohapatra:1980qe}%
  \BibitemOpen
  \bibfield  {author} {\bibinfo {author} {\bibfnamefont {R.~N.}\ \bibnamefont
  {Mohapatra}}\ and\ \bibinfo {author} {\bibfnamefont {R.~E.}\ \bibnamefont
  {Marshak}},\ }\href {\doibase 10.1103/PhysRevLett.44.1316} {\bibfield
  {journal} {\bibinfo  {journal} {Phys. Rev. Lett.}\ }\textbf {\bibinfo
  {volume} {44}},\ \bibinfo {pages} {1316} (\bibinfo {year} {1980})},\ \bibinfo
  {note} {[Erratum: Phys.Rev.Lett. 44, 1643 (1980)]}\BibitemShut {NoStop}%
\bibitem [{\citenamefont {Wetterich}(1981)}]{Wetterich:1981bx}%
  \BibitemOpen
  \bibfield  {author} {\bibinfo {author} {\bibfnamefont {C.}~\bibnamefont
  {Wetterich}},\ }\href {\doibase 10.1016/0550-3213(81)90279-0} {\bibfield
  {journal} {\bibinfo  {journal} {Nucl. Phys. B}\ }\textbf {\bibinfo {volume}
  {187}},\ \bibinfo {pages} {343} (\bibinfo {year} {1981})}\BibitemShut
  {NoStop}%
\bibitem [{\citenamefont {Dick}\ \emph {et~al.}(2000)\citenamefont {Dick},
  \citenamefont {Lindner}, \citenamefont {Ratz},\ and\ \citenamefont
  {Wright}}]{Dick:1999je}%
  \BibitemOpen
  \bibfield  {author} {\bibinfo {author} {\bibfnamefont {K.}~\bibnamefont
  {Dick}}, \bibinfo {author} {\bibfnamefont {M.}~\bibnamefont {Lindner}},
  \bibinfo {author} {\bibfnamefont {M.}~\bibnamefont {Ratz}}, \ and\ \bibinfo
  {author} {\bibfnamefont {D.}~\bibnamefont {Wright}},\ }\href {\doibase
  10.1103/PhysRevLett.84.4039} {\bibfield  {journal} {\bibinfo  {journal}
  {Phys. Rev. Lett.}\ }\textbf {\bibinfo {volume} {84}},\ \bibinfo {pages}
  {4039} (\bibinfo {year} {2000})},\ \Eprint
  {http://arxiv.org/abs/hep-ph/9907562} {arXiv:hep-ph/9907562} \BibitemShut
  {NoStop}%
\bibitem [{\citenamefont {Chao}(2016)}]{Chao:2015nsm}%
  \BibitemOpen
  \bibfield  {author} {\bibinfo {author} {\bibfnamefont {W.}~\bibnamefont
  {Chao}},\ }\href {\doibase 10.1103/PhysRevD.93.115013} {\bibfield  {journal}
  {\bibinfo  {journal} {Phys. Rev. D}\ }\textbf {\bibinfo {volume} {93}},\
  \bibinfo {pages} {115013} (\bibinfo {year} {2016})},\ \Eprint
  {http://arxiv.org/abs/1512.06297} {arXiv:1512.06297 [hep-ph]} \BibitemShut
  {NoStop}%
\bibitem [{\citenamefont {Chao}\ \emph {et~al.}(2017)\citenamefont {Chao},
  \citenamefont {Guo},\ and\ \citenamefont {Zhang}}]{Chao:2016avy}%
  \BibitemOpen
  \bibfield  {author} {\bibinfo {author} {\bibfnamefont {W.}~\bibnamefont
  {Chao}}, \bibinfo {author} {\bibfnamefont {H.-k.}\ \bibnamefont {Guo}}, \
  and\ \bibinfo {author} {\bibfnamefont {Y.}~\bibnamefont {Zhang}},\ }\href
  {\doibase 10.1007/JHEP04(2017)034} {\bibfield  {journal} {\bibinfo  {journal}
  {JHEP}\ }\textbf {\bibinfo {volume} {04}},\ \bibinfo {pages} {034} (\bibinfo
  {year} {2017})},\ \Eprint {http://arxiv.org/abs/1604.01771} {arXiv:1604.01771
  [hep-ph]} \BibitemShut {NoStop}%
\bibitem [{\citenamefont {He}\ \emph {et~al.}(1991)\citenamefont {He},
  \citenamefont {Joshi}, \citenamefont {Lew},\ and\ \citenamefont
  {Volkas}}]{He:1991qd}%
  \BibitemOpen
  \bibfield  {author} {\bibinfo {author} {\bibfnamefont {X.-G.}\ \bibnamefont
  {He}}, \bibinfo {author} {\bibfnamefont {G.~C.}\ \bibnamefont {Joshi}},
  \bibinfo {author} {\bibfnamefont {H.}~\bibnamefont {Lew}}, \ and\ \bibinfo
  {author} {\bibfnamefont {R.~R.}\ \bibnamefont {Volkas}},\ }\href {\doibase
  10.1103/PhysRevD.44.2118} {\bibfield  {journal} {\bibinfo  {journal} {Phys.
  Rev. D}\ }\textbf {\bibinfo {volume} {44}},\ \bibinfo {pages} {2118}
  (\bibinfo {year} {1991})}\BibitemShut {NoStop}%
\bibitem [{\citenamefont {Witten}(1982)}]{Witten:1982fp}%
  \BibitemOpen
  \bibfield  {author} {\bibinfo {author} {\bibfnamefont {E.}~\bibnamefont
  {Witten}},\ }\href {\doibase 10.1016/0370-2693(82)90728-6} {\bibfield
  {journal} {\bibinfo  {journal} {Phys. Lett. B}\ }\textbf {\bibinfo {volume}
  {117}},\ \bibinfo {pages} {324} (\bibinfo {year} {1982})}\BibitemShut
  {NoStop}%
\bibitem [{\citenamefont {Adler}(1969)}]{Adler:1969gk}%
  \BibitemOpen
  \bibfield  {author} {\bibinfo {author} {\bibfnamefont {S.~L.}\ \bibnamefont
  {Adler}},\ }\href {\doibase 10.1103/PhysRev.177.2426} {\bibfield  {journal}
  {\bibinfo  {journal} {Phys. Rev.}\ }\textbf {\bibinfo {volume} {177}},\
  \bibinfo {pages} {2426} (\bibinfo {year} {1969})}\BibitemShut {NoStop}%
\bibitem [{\citenamefont {Bell}\ and\ \citenamefont
  {Jackiw}(1969)}]{Bell:1969ts}%
  \BibitemOpen
  \bibfield  {author} {\bibinfo {author} {\bibfnamefont {J.~S.}\ \bibnamefont
  {Bell}}\ and\ \bibinfo {author} {\bibfnamefont {R.}~\bibnamefont {Jackiw}},\
  }\href {\doibase 10.1007/BF02823296} {\bibfield  {journal} {\bibinfo
  {journal} {Nuovo Cim. A}\ }\textbf {\bibinfo {volume} {60}},\ \bibinfo
  {pages} {47} (\bibinfo {year} {1969})}\BibitemShut {NoStop}%
\bibitem [{\citenamefont {Bardeen}(1969)}]{Bardeen:1969md}%
  \BibitemOpen
  \bibfield  {author} {\bibinfo {author} {\bibfnamefont {W.~A.}\ \bibnamefont
  {Bardeen}},\ }\href {\doibase 10.1103/PhysRev.184.1848} {\bibfield  {journal}
  {\bibinfo  {journal} {Phys. Rev.}\ }\textbf {\bibinfo {volume} {184}},\
  \bibinfo {pages} {1848} (\bibinfo {year} {1969})}\BibitemShut {NoStop}%
\bibitem [{\citenamefont {Eguchi}\ and\ \citenamefont
  {Freund}(1976)}]{Eguchi:1976db}%
  \BibitemOpen
  \bibfield  {author} {\bibinfo {author} {\bibfnamefont {T.}~\bibnamefont
  {Eguchi}}\ and\ \bibinfo {author} {\bibfnamefont {P.~G.~O.}\ \bibnamefont
  {Freund}},\ }\href {\doibase 10.1103/PhysRevLett.37.1251} {\bibfield
  {journal} {\bibinfo  {journal} {Phys. Rev. Lett.}\ }\textbf {\bibinfo
  {volume} {37}},\ \bibinfo {pages} {1251} (\bibinfo {year}
  {1976})}\BibitemShut {NoStop}%
\bibitem [{\citenamefont {Alvarez-Gaume}\ and\ \citenamefont
  {Witten}(1984)}]{Alvarez-Gaume:1983ihn}%
  \BibitemOpen
  \bibfield  {author} {\bibinfo {author} {\bibfnamefont {L.}~\bibnamefont
  {Alvarez-Gaume}}\ and\ \bibinfo {author} {\bibfnamefont {E.}~\bibnamefont
  {Witten}},\ }\href {\doibase 10.1016/0550-3213(84)90066-X} {\bibfield
  {journal} {\bibinfo  {journal} {Nucl. Phys. B}\ }\textbf {\bibinfo {volume}
  {234}},\ \bibinfo {pages} {269} (\bibinfo {year} {1984})}\BibitemShut
  {NoStop}%
\bibitem [{\citenamefont {Duerr}\ \emph {et~al.}(2013)\citenamefont {Duerr},
  \citenamefont {Fileviez~Perez},\ and\ \citenamefont {Wise}}]{Duerr:2013dza}%
  \BibitemOpen
  \bibfield  {author} {\bibinfo {author} {\bibfnamefont {M.}~\bibnamefont
  {Duerr}}, \bibinfo {author} {\bibfnamefont {P.}~\bibnamefont
  {Fileviez~Perez}}, \ and\ \bibinfo {author} {\bibfnamefont {M.~B.}\
  \bibnamefont {Wise}},\ }\href {\doibase 10.1103/PhysRevLett.110.231801}
  {\bibfield  {journal} {\bibinfo  {journal} {Phys. Rev. Lett.}\ }\textbf
  {\bibinfo {volume} {110}},\ \bibinfo {pages} {231801} (\bibinfo {year}
  {2013})},\ \Eprint {http://arxiv.org/abs/1304.0576} {arXiv:1304.0576
  [hep-ph]} \BibitemShut {NoStop}%
\bibitem [{\citenamefont {Turner}\ and\ \citenamefont
  {Widrow}(1988)}]{Turner:1987bw}%
  \BibitemOpen
  \bibfield  {author} {\bibinfo {author} {\bibfnamefont {M.~S.}\ \bibnamefont
  {Turner}}\ and\ \bibinfo {author} {\bibfnamefont {L.~M.}\ \bibnamefont
  {Widrow}},\ }\href {\doibase 10.1103/PhysRevD.37.2743} {\bibfield  {journal}
  {\bibinfo  {journal} {Phys. Rev. D}\ }\textbf {\bibinfo {volume} {37}},\
  \bibinfo {pages} {2743} (\bibinfo {year} {1988})}\BibitemShut {NoStop}%
\bibitem [{\citenamefont {Garretson}\ \emph {et~al.}(1992)\citenamefont
  {Garretson}, \citenamefont {Field},\ and\ \citenamefont
  {Carroll}}]{Garretson:1992vt}%
  \BibitemOpen
  \bibfield  {author} {\bibinfo {author} {\bibfnamefont {W.~D.}\ \bibnamefont
  {Garretson}}, \bibinfo {author} {\bibfnamefont {G.~B.}\ \bibnamefont
  {Field}}, \ and\ \bibinfo {author} {\bibfnamefont {S.~M.}\ \bibnamefont
  {Carroll}},\ }\href {\doibase 10.1103/PhysRevD.46.5346} {\bibfield  {journal}
  {\bibinfo  {journal} {Phys. Rev. D}\ }\textbf {\bibinfo {volume} {46}},\
  \bibinfo {pages} {5346} (\bibinfo {year} {1992})},\ \Eprint
  {http://arxiv.org/abs/hep-ph/9209238} {arXiv:hep-ph/9209238} \BibitemShut
  {NoStop}%
\bibitem [{\citenamefont {Anber}\ and\ \citenamefont
  {Sorbo}(2006)}]{Anber:2006xt}%
  \BibitemOpen
  \bibfield  {author} {\bibinfo {author} {\bibfnamefont {M.~M.}\ \bibnamefont
  {Anber}}\ and\ \bibinfo {author} {\bibfnamefont {L.}~\bibnamefont {Sorbo}},\
  }\href {\doibase 10.1088/1475-7516/2006/10/018} {\bibfield  {journal}
  {\bibinfo  {journal} {JCAP}\ }\textbf {\bibinfo {volume} {10}},\ \bibinfo
  {pages} {018} (\bibinfo {year} {2006})},\ \Eprint
  {http://arxiv.org/abs/astro-ph/0606534} {arXiv:astro-ph/0606534} \BibitemShut
  {NoStop}%
\bibitem [{\citenamefont {Weinberg}(2008)}]{Weinberg:2008zzc}%
  \BibitemOpen
  \bibfield  {author} {\bibinfo {author} {\bibfnamefont {S.}~\bibnamefont
  {Weinberg}},\ }\href@noop {} {\emph {\bibinfo {title} {{Cosmology}}}}\
  (\bibinfo {year} {2008})\BibitemShut {NoStop}%
\bibitem [{\citenamefont {Maleknejad}(2016)}]{Maleknejad:2016qjz}%
  \BibitemOpen
  \bibfield  {author} {\bibinfo {author} {\bibfnamefont {A.}~\bibnamefont
  {Maleknejad}},\ }\href {\doibase 10.1007/JHEP07(2016)104} {\bibfield
  {journal} {\bibinfo  {journal} {JHEP}\ }\textbf {\bibinfo {volume} {07}},\
  \bibinfo {pages} {104} (\bibinfo {year} {2016})},\ \Eprint
  {http://arxiv.org/abs/1604.03327} {arXiv:1604.03327 [hep-ph]} \BibitemShut
  {NoStop}%
\bibitem [{\citenamefont {Domcke}\ and\ \citenamefont
  {Mukaida}(2018)}]{Domcke:2018eki}%
  \BibitemOpen
  \bibfield  {author} {\bibinfo {author} {\bibfnamefont {V.}~\bibnamefont
  {Domcke}}\ and\ \bibinfo {author} {\bibfnamefont {K.}~\bibnamefont
  {Mukaida}},\ }\href {\doibase 10.1088/1475-7516/2018/11/020} {\bibfield
  {journal} {\bibinfo  {journal} {JCAP}\ }\textbf {\bibinfo {volume} {11}},\
  \bibinfo {pages} {020} (\bibinfo {year} {2018})},\ \Eprint
  {http://arxiv.org/abs/1806.08769} {arXiv:1806.08769 [hep-ph]} \BibitemShut
  {NoStop}%
\bibitem [{\citenamefont {Maleknejad}\ and\ \citenamefont
  {Komatsu}(2019)}]{Maleknejad:2018nxz}%
  \BibitemOpen
  \bibfield  {author} {\bibinfo {author} {\bibfnamefont {A.}~\bibnamefont
  {Maleknejad}}\ and\ \bibinfo {author} {\bibfnamefont {E.}~\bibnamefont
  {Komatsu}},\ }\href {\doibase 10.1007/JHEP05(2019)174} {\bibfield  {journal}
  {\bibinfo  {journal} {JHEP}\ }\textbf {\bibinfo {volume} {05}},\ \bibinfo
  {pages} {174} (\bibinfo {year} {2019})},\ \Eprint
  {http://arxiv.org/abs/1808.09076} {arXiv:1808.09076 [hep-ph]} \BibitemShut
  {NoStop}%
\bibitem [{\citenamefont {Adshead}\ \emph {et~al.}(2018)\citenamefont
  {Adshead}, \citenamefont {Pearce}, \citenamefont {Peloso}, \citenamefont
  {Roberts},\ and\ \citenamefont {Sorbo}}]{Adshead:2018oaa}%
  \BibitemOpen
  \bibfield  {author} {\bibinfo {author} {\bibfnamefont {P.}~\bibnamefont
  {Adshead}}, \bibinfo {author} {\bibfnamefont {L.}~\bibnamefont {Pearce}},
  \bibinfo {author} {\bibfnamefont {M.}~\bibnamefont {Peloso}}, \bibinfo
  {author} {\bibfnamefont {M.~A.}\ \bibnamefont {Roberts}}, \ and\ \bibinfo
  {author} {\bibfnamefont {L.}~\bibnamefont {Sorbo}},\ }\href {\doibase
  10.1088/1475-7516/2018/06/020} {\bibfield  {journal} {\bibinfo  {journal}
  {JCAP}\ }\textbf {\bibinfo {volume} {06}},\ \bibinfo {pages} {020} (\bibinfo
  {year} {2018})},\ \Eprint {http://arxiv.org/abs/1803.04501} {arXiv:1803.04501
  [astro-ph.CO]} \BibitemShut {NoStop}%
\bibitem [{\citenamefont {Kamada}\ and\ \citenamefont
  {Long}(2016)}]{Kamada:2016cnb}%
  \BibitemOpen
  \bibfield  {author} {\bibinfo {author} {\bibfnamefont {K.}~\bibnamefont
  {Kamada}}\ and\ \bibinfo {author} {\bibfnamefont {A.~J.}\ \bibnamefont
  {Long}},\ }\href {\doibase 10.1103/PhysRevD.94.123509} {\bibfield  {journal}
  {\bibinfo  {journal} {Phys. Rev. D}\ }\textbf {\bibinfo {volume} {94}},\
  \bibinfo {pages} {123509} (\bibinfo {year} {2016})},\ \Eprint
  {http://arxiv.org/abs/1610.03074} {arXiv:1610.03074 [hep-ph]} \BibitemShut
  {NoStop}%
\end{thebibliography}%

\end{document}